\documentclass[usenatbib, useAMS]{mnras}

\usepackage{graphicx}
\usepackage{amsmath,paralist,xcolor,xspace,amssymb}
\usepackage{times}

\newcommand{\msun}{{\rm{M}_\odot}}
\newcommand{\HI}{\ion{H}{I}\xspace}
\newcommand{\rhocr}{\rho_{\rm{cr}}}

\newcommand{\lcdm}{$\Lambda$CDM\xspace}

\newcommand{\eagle}{\textsc{eagle}\xspace}

\newcommand{\gadget}{\textsc{gadget}\xspace} 
\newcommand{\subfind}{\textsc{subfind}\xspace}
\newcommand{\anarchy}{\textsc{anarchy}\xspace}

\title[EAGLE: Hydrodynamics scheme variations]{The EAGLE simulations of galaxy
  formation: the importance of the hydrodynamics scheme}

\author[M. Schaller et al.]  {Matthieu Schaller$^1$\thanks{E-mail: matthieu.schaller@durham.ac.uk},
                             Claudio Dalla Vecchia$^{2,3}$,
                             Joop Schaye$^{4}$,
			     Richard G. Bower$^{1}$, \newauthor
                             Tom Theuns$^{1}$,
                             Robert A. Crain$^{5}$,
                             Michelle Furlong$^{1}$ \&
                             Ian G. McCarthy$^{5}$
                             \\
$^1$Institute for Computational Cosmology, Durham University, South Road, Durham, UK, DH1 3LE\\
$^2$Instituto de Astrof\'isica de Canarias, C/ V\'ia L\'actea s/n, 38205 La Laguna, Tenerife, Spain\\
$^3$Departamento de Astrof\'isica, Universidad de La Laguna, Av. del Astrof\'isico Franciso S\'anchez s/n, 38206 
La Laguna, Tenerife, Spain\\
$^4$Leiden Observatory, Leiden University, P.O. Box 9513, 2300 RA Leiden, The Netherlands\\
$^5$Astrophysics Research Institute, Liverpool John Moores University, 146 Brownlow Hill, Liverpool L3 5RF, UK
}
\begin{document}

\date{\today}

\pagerange{\pageref{firstpage}--\pageref{lastpage}} \pubyear{2015}

\maketitle

\label{firstpage}

\begin{abstract}
We present results from a subset of simulations from the ``Evolution and
Assembly of GaLaxies and their Environments'' (\eagle) suite in which the
formulation of the hydrodynamics scheme is varied. We compare simulations that
use the same subgrid models without re-calibration of the parameters but
employing the standard \gadget flavour of smoothed particle hydrodynamics (SPH)
instead of the more recent state-of-the-art \anarchy formulation of SPH that was
used in the fiducial \eagle runs. We find that the properties of most galaxies,
including their masses and sizes, are not significantly affected by the details
of the hydrodynamics solver.  However, the star formation rates of the most
massive objects are affected by the lack of phase mixing due to spurious surface
tension in the simulation using standard SPH.  This affects the efficiency with
which AGN activity can quench star formation in these galaxies and it also leads
to differences in the intragroup medium that affect the X-ray emission from
these objects. The differences that can be attributed to the hydrodynamics
solver are, however, likely to be less important at lower resolution. We also
find that the use of a time step limiter is important for achieving the feedback
efficiency required to match observations of the low-mass end of the galaxy
stellar mass function.
\end{abstract}


\section{Introduction}
\label{sec:introduction}

Cosmological hydrodynamical simulations have started to play a major role in the
study of galaxy formation. Recent simulations are able to cover the large dynamical
range required to study the large-scale structure dominated by dark matter as well as
the centres of halos where baryon physics dominates the evolution. Comparisons of
such simulations with observations show broad agreement and help confirm the
predictions of the $\Lambda$CDM paradigm \citep[e.g.][]{Vogelsberger2014,Schaye2015}.

Galaxy formation involves a mixture of complex processes and the numerical
requirements to simulate all of the relevant scales are enormous. A direct
consequence of this is the need to model some of the unresolved processes with
subgrid prescriptions. Other processes, taking place on larger scales, can in
principle be followed accurately by numerical hydrodynamics solvers. The shocking of
cold gas penetrating halos and the turbulence generated by supernova activity within
galaxies are examples of the processes that can, in principle, be treated by the
hydrodynamics solver. Conversely, the accretion of gas onto black holes and the
formation and evolution of stars are examples of processes that occur on scales that
are too small to be simulated jointly with the large-scale environment. In practice,
however, these two categories of processes are interleaved and it is hence difficult to
demonstrate convergence even of the purely hydrodynamical processes. Practitioners are
therefore forced to chose a numerical hydrodynamics solver that gives accurate results at
the resolution of interest.

Many numerical techniques (e.g. Adaptive Mesh Refinement, particle techniques, moving-mesh
techniques, mesh-free techniques)  have been developed over the years to solve
the equations of hydrodynamics, each of them coming in different ``flavours'',
i.e. coming with slightly different equations, assumptions and limitations.  For the
processes that can be simulated using standard numerical solvers, the main question
is how the various parameters that enter these hydrodynamics solvers affect the
formation of galaxies in the simulations. For example, it has been reported that
different numerical techniques and choices of parameters affect the disruption of a
cold gas blob in a low-density hot medium, a case directly relevant the accretion of
gas and satellite of galaxies \citep{Frenk1999, Marri2003, Okamoto2003, OShea2005,
  Agertz2007, Wadsley2008, Mitchell2009, Keres2012, Sijacki2012}. In principle, the
values of these numerical parameters can be set by performing controlled numerical
experiments for which the solution is known.

In the case of simulations using Smoothed Particle Hydrodynamics (SPH) solvers
\citep{Lucy1977, Gingold1977} \citep[see][for reviews]{Springel2010b,Price2012}, the
free parameters relate to the treatment of shocks, artificial viscosity and
conduction and are related to the way the SPH equations are derived, leading to
different flavours of the technique. Performing controlled tests such as Sedov
explosions or Kelvin-Helmoltz instabilities \citep[e.g.][]{Price2008, Read2010,
  Springel2010a, Hu2014, Hopkins2015, Beck2015} enables the simulator to identify
well-motivated values for the parameters and understand the limitations of the
formulation. Early flavours of SPH had issues dealing with discontinuities in the
fluid. One of many examples of this problem is the ``blob test'' of \cite{Agertz2007}
which was widely used in the literature to demonstrate the failure of SPH. A lot of
effort has then been spent by the community to improve the situation and many
alternative solutions have been proposed to overcome the appearance of spurious
surface tension that prevents the correct mixing of phases. Solutions using either an
alternative formulation of the equations in which the discontinuities are smoothed
were proposed \citep[e.g.][]{Ritchie2001, Read2010, Abel2011, Saitoh2013} as well as
solution involving additional terms diffusing material across the discontinuities
\citep[e.g.][]{Price2008}. Both types of solutions to the discontinuity problem
present shortcomings (see discussion at the end of Section \ref{ssec:SPH}) and this
motivated the implementation of SPH used in this paper, which uses a combination of
both solutions. Although one can in principle calibrate the free parameters using
tests, it is unclear whether there is a single set of values that is suitable for all
problems and whether these parameter values are also the best choice when performing
simulations of very hot and diffuse conditions, such as those present in the hot
haloes of galaxies \citep[e.g.][]{Sembolini2015}. Moreover, the large gap in
resolution between these controlled experiments and cosmological simulations makes
the extrapolation of the solver's behaviour a difficult and uncertain task. The
correct treatment of entropy jumps across shocks or of the spurious viscosity that
can appear in differentially rotating disks can have direct consequences for the
population of simulated galaxies.

In their comprehensive study of galaxy formation models, \cite{Scannapieco2012} used
multiple hydrodynamics solvers coupled to multiple sets of subgrid models to simulate
the formation of galaxies in a single halo and to study the relative impact of the
choice of solver and subgrid model. One of their main findings was that the
variations in the hydrodynamics solvers led to much smaller changes in the final
results than did the changes to the subgrid model parameters. This was especially the
case for the prescription of feedback, which can change the final galaxy tremendously
\citep[e.g.][]{Schaye2010, Haas2013, Vogelsberger2013,Crain2015}. A more controlled
experiment was performed by \cite{Keres2012}, who compared two hydrodynamics solvers
but used only a simplified model of galaxy formation and, apart for the most massive
galaxies, found very little difference in the galaxy population despite the large gap
in accuracy between the hydro solvers tested. Their two simulations, however,
displayed significant differences in the gas properties, especially in the cold gas
fractions. More realistic subgrid models, especially of feedback, are likely to
suppress some of these difference.

Building on those studies, we attempt to quantify the impact of the uncertainties in
two different implementations of SPH solvers on a simulated galaxy population. The
\eagle simulation project \citep{Schaye2015, Crain2015} uses a state-of-the-art
implementation of SPH, called \anarchy \citep[Dalla Vecchia in prep., see also
  appendix A of][]{Schaye2015} and the time-step limiter of
\cite{Durier2012}. \eagle's subgrid model parameters were calibrated to reproduce the
observed local Universe population of galaxies. In this study, we vary the
hydrodynamics solver. We compare \anarchy to the older \cite{Springel2002} flavour of SPH
implemented in the \gadget code \citep{Springel2005} and compare the resulting galaxy
population to the one in the reference \eagle simulation and to those in simulations
with weaker/stronger stellar feedback and to runs without AGN feedback. Since \eagle
broadly reproduces the observed galaxy population, our test is especially relevant
and enables us to disentangle the effects of the hydro solver from the effects of the
subgrid model.

This paper is structured as follows. In section \ref{sec:simulations} the \eagle
model and the two flavours of SPH that we consider are described. Section
\ref{sec:galaxyPopulation} discusses the impact of the hydrodynamics solver on the
simulated galaxies whilst Section \ref{sec:gasProperties} presents differences in the
gas properties of the haloes. A summary of our findings can be found in Section
\ref{sec:summary}.

Throughout this paper, we assume a \emph{Planck2013} flat \lcdm cosmology
\citep{Planck2013} ($h = 0.6777$, $\Omega_b = 0.04825$, $\Omega_m = 0.307$ and
$\sigma_8= 0.8288$) and express all quantities without $h$ factors.

\section{The EAGLE simulations}
\label{sec:simulations}

The \eagle set consists of a series of cosmological simulations with state-of-the-art
subgrid models and smoothed particle hydrodynamics. The simulations have been
calibrated to reproduce the observed galaxy stellar mass function, the relation
between galaxy stellar mass and supermassive black hole mass and galaxy mass-size
relation at $z=0.1$. The simulations also broadly reproduce a large variety of other
observables such as the Tully-Fisher relation and specific star formation rates
\citep{Schaye2015}, the $\rm{H}_2$ and \ion{H}{I} properties of galaxies \citep[][
  Bahe et al. submitted]{Lagos2015}, the evolution of the galaxy stellar mass
function \citep{Furlong2015}, the column density distribution of intergalactic metals
\citep{Schaye2015} and \HI \citep{Rahmati2015} as well as galaxy rotation curves
\citep{Schaller2015} and luminosities \citep{trayford2015}.

The \eagle simulations discussed in this paper follow $752^3\approx4.3\times10^8$
dark matter particles and the same number of gas particles in a $50^3~\rm{Mpc}^3$
cubic volume from $\Lambda$CDM initial conditions. Note that the simulation volumes
considered here are a factor of eight smaller than the main $100^3~\rm{Mpc}^3$ \eagle
run. The mass of a dark matter particle is $m_{\rm DM}=9.7\times10^6~\msun$ and the
initial mass of a gas particle is $m_{\rm g}=1.8\times10^6~\msun$. The gravitational
softening length is $700~\rm{pc}$ (Plummer equivalent) in physical units below
$z=2.8$ and $2.66~\rm{kpc}$ (comoving) at higher redshifts. The simulations were run
with a heavily modified version of the \gadget-3 $N$-body tree-PM and SPH code, last
described in \cite{Springel2005}. The changes include the introduction of the subgrid
models described in the next subsection as well as the implementation of the \anarchy
flavour of SPH, whose impact on the simulation outcome is the topic of this paper. In
the next subsections we will describe the subgrid model used in the \eagle
simulations with a special emphasis on those aspects of the model that are directly
impacted upon by the hydrodynamic scheme. For the sake of completeness, we then
briefly describe both the standard \gadget and \anarchy flavours of SPH.

\subsection{Subgrid models and halo identification}
\label{ssec:subgrid}

Radiative cooling is implemented using element-by-element rates \citep{Wiersma2009a}
for the $11$ most important metals in the presence of the CMB and UV/X-ray
backgrounds given by \cite{Haardt2001}. To prevent artificial fragmentation, the cold
and dense gas is not allowed to cool to temperatures below those corresponding to an
equation of state $P_{\rm eos} \propto \rho^{4/3}$ that is designed to keep the Jeans
mass marginally resolved \citep{Schaye2008}. Star formation is implemented using a
pressure-dependent prescription that reproduces the observed Kennicutt-Schmidt star
formation law \citep{Schaye2008} and uses a threshold that captures the metallicity
dependence of the transition from the warm, atomic to the cold, molecular gas phase
\citep{Schaye2004}.  Star particles are treated as single stellar populations with a
\cite{Chabrier2003} IMF evolving along the tracks provided by
\cite{Portinari1998}. Metals from supernovae and AGB stars are injected into the
interstellar medium (ISM) following the model of \cite{Wiersma2009b} and stellar
feedback is implemented by the stochastic injection of thermal energy into the gas as
described in \cite{DallaVecchia2012}. The amount of energy injected into the ISM per
feedback event depends on the local gas metallicity and density in an attempt to take
into account the unresolved structure of the ISM \citep{Schaye2015, Crain2015}.
Supermassive black hole seeds are injected in halos above $10^{10}h^{-1}\msun$ and
grow through mergers and accretion of low angular momentum gas
\citep{RosasGuevara2013,Schaye2015}. AGN feedback is performed by injecting thermal
energy into the gas directly surrounding the black hole
\citep{Booth2009,DallaVecchia2012}.

The subgrid model was calibrated (by adjusting the intensity of stellar feedback and
the accretion rate onto black holes) so as to reproduce the present-day galaxy
stellar mass function and galaxy sizes \citep{Schaye2015}. As discussed by
\cite{Crain2015}, this latter requirement is crucial to obtain a galaxy population
that evolves with redshift in a similar fashion to the observed populations
\citep{Furlong2015}.

Haloes were identified using the Friends-of-Friends (FoF) algorithm \citep{Davis1985}
with linking length $0.2$ times the mean interparticle distance, and bound structures
within them were then identified using the \subfind code \citep{Springel2001,
  Dolag2009}. A sphere centred at the minimum of the gravitational potential of each
subhalo is grown until the mass contained within a given radius, $R_{200}$, reaches
$M_{200} = 200\left(4\upi\rhocr(z)R_{200}^3/3\right)$, where $\rhocr(z)=3H(z)^2/8\upi
G$ is the critical density at the redshift of interest.

\subsection{SPH implementations}
\label{ssec:SPH}

All simulations that are compared in this study use modifications of the Gadget-3
code. We use both the default flavour of SPH documented in \cite{Springel2005} and
the more recent flavour nicknamed \anarchy \citep[Dalla Vecchia (in prep.), see also
  appendix A of][]{Schaye2015} implemented as a modification to the default code. For
completeness, we describe both sets of hydrodynamical equations in this section
without derivations. For comprehensive descriptions and motivations, see the review
by \cite{Price2012} and the description of the alternative formalism by
\cite{Hopkins2013,Hopkins2015}. A formulation of SPH that is similar to \anarchy is
presented in \cite{Hu2014}.  Note that apart from the differences highlighted in this
section, the codes (and parameters) used for both types of simulations are identical.

\subsubsection{Default \gadget-2 SPH}

In its default version, \gadget-2 uses the fully conservative SPH equations
introduced by \cite{Springel2002}. We will label this ``\gadget SPH'' in the
remainder of this paper and restrict our discussion of the model to the 3D case. As
in any flavour of SPH, the starting point is a the choice of a smoothing function to
reconstruct field quantities at any point in space from a weighted average over the
surrounding particles. In the case of gas density, at position $\mathbf{x}_i$. the
equation reads

\begin{equation}
 \rho_i = \sum_j m_j W\left(|\mathbf{x}_i - \mathbf{x}_j|, h_i\right),
 \label{eq:gadget_density}
\end{equation}
where $W(|\mathbf{r}|,h)$ is the spherically symmetric kernel function. In the
case of \gadget, the $M_4$ cubic B-spline function is used and reads

\begin{equation}
 W(r,h) = \frac{8}{\upi h^3}\left\lbrace
 \begin{array}{rcl}
  1 - 6\left(\frac{r}{h}\right)^2 + 6\left(\frac{r}{h}\right)^3 & \mathrm{if} &
  0\leq r\leq\frac{h}{2}\\ 2\left(1-\frac{r}{h}\right)^3 & \mathrm{if} &
  \frac{h}{2} < r\leq h\\ 0 & \mathrm{if} & r > h.
 \end{array}
\right.\nonumber
\end{equation}

The smoothing length $h_i$ of a particle is obtained by requiring that the weighted
number of neighbours
\begin{equation}
 N_{\rm{ngb}} = \frac{4}{3}\upi h_i^3\sum_j W\left(|\mathbf{x}_i - \mathbf{x}_j|,
 h_i\right)
\end{equation}
of the particle is close to a pre-defined constant; $N_{\rm ngb}=48$ in our
case. Note, however, that contrary to what is often written in the literature, \gadget
defines the smoothing length as the cut-off radius of the kernel and not as the more
physical FWHM of the kernel function \citep{Dehnen2012}.

The quantity integrated in time alongside the velocities and positions of the
particles is the entropic function\footnote{This quantity is not the thermodynamic
  entropy $s$ but a monotonic function of it.}  $A_i=P_i/\rho_i^\gamma$, defined in
terms of the pressure $P_i$ and polytropic index $\gamma$. The equations of motion
are then given by

\begin{equation}
 \frac{\rm{d}\mathbf{v}_i}{\rm{d}t} = -\sum_j
 m_j\left[\frac{P_i}{\Omega_i\rho_i^2}
   \nabla_iW_{ij}(h_i)+\frac{P_j}{\Omega_j\rho_j^2}\nabla_iW_{ij}(h_j)\right],
\label{eq:gadget_eom}
\end{equation}
where $\Omega_i$ accounts for the gradient of the smoothing length,

\begin{equation}
 \Omega_i = 1 + \frac{h_i}{3\rho_i}\sum_j m_j\frac{\partial
   W_{ij}(h_i)}{\partial h}
\end{equation}
and $W_{ij}(h_i) \equiv W(\mathbf{x}_i - \mathbf{x}_j,h_i)$. In the absence of
radiative cooling or thermal diffusion terms, the entropic function of each particle
is a constant in time. Only radiative cooling, feedback events (see the previous section) and
shocks will change the entropic function.

In order to capture shocks, artificial viscosity is implemented by adding a term to
the equations of motion (eq. \ref{eq:gadget_eom}) to evolve the entropic function
accordingly:

\begin{eqnarray}
 \frac{\rm{d}\mathbf{v}_i}{\rm{d}t} &\stackrel{\rm{visc.}}{=}&
 -\frac{1}{4}\sum_jm_j \Pi_{ij} \nabla \overline{W}_{ij} \left(f_i + f_j\right)
 \nonumber
 \\ 
\frac{\rm{d}A_i}{\rm{d}t}&\stackrel{\rm{visc.}}{=}&\frac{1}{8}\frac{\gamma-1}{
\rho_i^{\gamma-1}}
 \sum_j m_j \Pi_{ij} \left(\mathbf{v}_i-\mathbf{v}_j\right)\cdot\nabla
 \overline{W}_{ij}\left(f_i + f_j\right), \nonumber
\end{eqnarray}
with $\overline{W}_{ij} \equiv \left(W_{ij}(h_i) + W_{ij}(h_j)\right)$ and the viscous
tensor ($\Pi_{ij}$) and shear flow switch $f_i$ defined below. Following
\cite{Monaghan1997}, the viscous tensor, which plays the role of an additional
pressure in the equations of motion, is defined in terms of the particle's sound
speed, $c_i=\sqrt{\gamma P_i/\rho_i}$, as

\begin{eqnarray}
 \Pi_{ij} &=& -\alpha\frac{\left(c_i + c_j - 3w_{ij}\right)w_{ij}}{\rho_i +
   \rho_j}, \label{eq:P_ij}\\ w_{ij} &=& \min\left(0, 
\frac{\left(\mathbf{v}_j-\mathbf{v}_i\right)\cdot\left(\mathbf{x}_i-\mathbf{x}
_j\right)}{\left|\mathbf{x}_i-\mathbf{x}_j
   \right|} \right) \label{eq:w_ij}
\end{eqnarray}
with the dimensionless viscosity parameter set to the commonly used value of
$\alpha=2$ in our simulations. Finally, to prevent the application of viscosity
in the case of pure shear flows, the switch proposed by \cite{Balsara1995} is
used:

\begin{equation}
 f_i 
=\frac{\left|\nabla\cdot\mathbf{v}_i\right|}{\left|\nabla\cdot\mathbf{v}
_i\right|
   + \left|\nabla\times\mathbf{v}_i\right| + 10^{-4}c_i/h_i},
\end{equation}
with the last term in the denominator added to avoid numerical instabilities. The
divergence and curl of the velocity field are computed in the standard SPH way
\citep[e.g.][]{Price2012}.

\subsubsection{\anarchy SPH}

The first change in \anarchy with respect to \gadget is the choice of kernel
function.  More accurate estimators for both the field quantities and their
derivatives can be obtained by using \cite{Wendland1995} kernels
\citep{Dehnen2012}. \anarchy uses the $C_2$ kernel. This kernel function is not
affected by the pairing instability, which occurs when high values of $N_{\rm ngb}$
are used with spline kernels. It reads

\begin{equation}
 W(r,h) = \frac{21}{2\upi h^3} \left\lbrace \begin{array}{rcl}
   \left(1-\frac{r}{h}\right)^4\left(1+4\frac{r}{h}\right) & \mathrm{if} & 0\leq
   r\leq h\\ 0 & \mathrm{if} & r > h.
 \end{array}
\right.\nonumber
\end{equation}

To keep the effective resolution of the simulation similar between the two flavours
of SPH, we use $N_{\rm ngb}=58$ with this kernel. This yields the same kernel FWHM as
obtained for the cubic kernel\footnote{Expressing our resolution in terms of the
  local inter-particle separation \citep{Price2012,Dehnen2012} gives
  $\eta={\rm{FWHM}}(W(r,h))/\Delta x=1.235$ for both kernels.} with $N_{\rm
  ngb}=48$. Note, however, that the $C_2$ kernel only exhibits better behaviour than
the cubic spline kernel when large numbers of neighbours ($N_{\rm ngb}\gtrsim100$) are
used \citep{Dehnen2012}. We use the $C_2$ kernel with $N_{\rm ngb}=58$ to be
consistent with both the \eagle resolution and the hydrodynamics studies of Dalla
Vecchia (in prep.) who used the same kernel but more neighbours.

The equations of motion used in the \anarchy flavour of SPH are based on the
Pressure-Entropy formulation of \cite{Hopkins2013}, a generalisation of the earlier
solutions of \cite{Ritchie2001}, \cite{Read2010}, \cite{Abel2011} and \cite{Saitoh2013}. The two
quantities carried by particles that are integrated forward in time are again the
velocity and the entropic function.  Alongside the density, which is computed in the
usual way (Eq. \ref{eq:gadget_density}), two additional smoothed quantities are
introduced in this formulation of SPH: the weighted density

\begin{equation}
 \bar\rho_i = \frac{1}{A_i^{1/\gamma}}\sum_{j}m_j
 A_j^{1/\gamma}W\left(|\mathbf{x}_i - \mathbf{x}_j|, h_i\right)
\end{equation}
and its associated weighted pressure, $\bar P_i = A_i \bar\rho_i^\gamma$. Despite
having the same units as the regular density, its weighted counterpart should only be
understood as an intermediate quantity entering other equations and should not be
used as the gas density. Using these two new quantities, the equation of motion for
the particle velocities becomes

\begin{eqnarray}
  \frac{\rm{d}\mathbf{v}_i}{\rm{d}t} &=& -\sum_j
  m_j\left[\frac{A_j^{1/\gamma}}{A_i^{1/\gamma}}\frac{\bar P_i}{\bar
      \rho_i^2}\Omega_{ij}\nabla_iW_{ij}(h_i)~+ \right.  \nonumber\\ & &
    \left.\frac{A_i^{1/\gamma}}{A_j^{1/\gamma}}\frac{\bar P_j}{\bar
      \rho_j^2}\Omega_{ji}\nabla_jW_{ij}(h_j) \right]
  \label{eq:anarchy_eom}
\end{eqnarray}
with the terms accounting for the gradients in the smoothing length reading

\begin{equation}
 \Omega_{ij} = 1 -
 \frac{1}{A_i^{1/\gamma}}\left(\frac{h_i}{3\rho_i}\frac{\partial \bar
   P_i^{1/\gamma}}{\partial
   h_i}\right)\left(1+\frac{h_i}{3\rho_i}\frac{}{}\right)^{-1}.
\end{equation}

The use of the smoothed quantities $\bar P_i$ and $\bar\rho_i$ in the equations of
motion smooths out the spurious pressure jumps appearing at contact discontinuities
in older formulations of SPH \citep{Saitoh2013, Hopkins2013}.

As in all versions of SPH, artificial viscosity has to be added to capture shocks. In
the \anarchy formulation of SPH, this is done following the method of
\cite{Cullen2010}. Their scheme is the latest iteration of a series of improvements
to the standard \citep{Monaghan1997} viscosity term that started with the proposal of
\cite{Morris1997} to assign individual viscosities $\alpha_i$ to each
particle. Improving on the work of \cite{Rosswog2000}, \cite{Price2004} and
\cite{Wetzstein2009}, \cite{Cullen2010} proposed a differential equation for
$\alpha_i$ that is solved alongside the equations of motion
(Eqn. \ref{eq:anarchy_eom}):

\begin{equation}
\dot\alpha_i = 2lv_{{\rm sig},i}\left( \alpha_{{\rm loc},i} - \alpha_i\right) /
h_i,
\end{equation}
with $l=0.01$ and the signal velocity $v_{{\rm sig},i}$ introduced below. The
local viscosity estimator $\alpha_{{\rm loc},i}$ is given by

\begin{equation}
 \alpha_{{\rm loc},i} = \alpha_{\rm max} \frac{h_i^2S_i}{v_{{\rm sig},i}^2 +
   h_i^2S_i},
\end{equation}
where $\alpha_{\rm max} = 2$ and $S_i = \max\left(0, -\frac{d}{d
  t}\left(\nabla\cdot\mathbf{v}_i\right)\right)$ is the shock detector. After passing
through a shock, $S_i=0$ and hence $\alpha_{{\rm loc},i}=0$, leading to a decrease in
$\alpha_i$. We impose $\alpha_i > \alpha_{\rm min}=0.05$ to facilitate particle
re-ordering. The signal velocity is constructed to capture the maximum velocity at
which information can be transferred between particles whilst remaining positive:

\begin{equation}
 v_{{\rm sig},i} = \max_{|\mathbf{x}_{ij}|\leq
   h_i}\left(\frac{1}{2}\left(c_i+c_j\right)-\min(0,
 \mathbf{v}_{ij}\cdot\hat{\mathbf{x}}_{ij})\right),
\end{equation}
with $\hat{\mathbf{x}}_{ij} = (\mathbf{x}_i - \mathbf{x}_j) / |\mathbf{x}_i -
\mathbf{x}_j|$ and $\mathbf{v}_{ij} = \mathbf{v}_j - \mathbf{v}_i$.

The individual viscosity coefficients $\alpha_i$ are then combined to enter the
equations of motion in a similar way as in the \gadget formulation. Equations
\ref{eq:P_ij} and \ref{eq:w_ij} are replaced by:

\begin{eqnarray}
 \Pi_{ij} &=& -\frac{\alpha_i+\alpha_j}{2}\frac{\left(c_i + c_j -
   3w_{ij}\right)w_{ij}}{\rho_i + \rho_j}, \\ w_{ij} &=& \min\left(0, 
\frac{\left(\mathbf{v}_j-\mathbf{v}_i\right)\cdot\left(\mathbf{x}_i-\mathbf{x}
_j\right)}{\left|\mathbf{x}_i-\mathbf{x}_j
   \right|} \right).
\end{eqnarray}

Note that contrary to \cite{Hu2014}, we do not implement expensive matrix calculations
\citep{Cullen2010} for the calculation of the velocity divergence time derivative
entering the shock detector $S_i$ as we found that using the standard SPH expressions
was sufficient for the accuracy we targeted.

The last improvement included in the \anarchy flavour of SPH is the use of some
entropy diffusion between particles.  SPH is by construction non-diffusive
\citep[e.g.][]{Price2012} and does, hence, not incorporate the thermal conduction
that may be required to faithfully reproduce the micro-scale mixing of gas phases. We
implement a small level of numerical diffusion following the recipe of
\cite{Monaghan1997} and \cite{Price2008}. We compute the internal energies from the
entropies and these are then used in the equations for the diffusion. The use of the
pressure-entropy formalism (eq. \ref{eq:anarchy_eom}) prevents the formation of
spurious surface tension at contact discontinuities \citep{Hopkins2013}. This small
amount of numerical diffusion allows the particles to mix their entropies at the
discontinuity and hence create one single phase (Dalla Vecchia in prep.). The
diffusion is hence used to solve a numerical problem and not to introduce a
macroscopic conduction.  This results in cluster entropy profiles in agreement with
the results from grid and moving-mesh codes \citep[see the comparison of][which
  includes \anarchy]{Sembolini2015}.  We compute the rate of change of the conduction
using the second derivative of the energy. This means that large conduction values
$\alpha_{\rm diff}$ are triggered by discontinuities in the first derivative of the
energy, not by smooth pressure gradients as in (self-)gravitating objects.  Moreover,
$\alpha_{\rm diff}$ may take some time to increase while the smoothing of the
discontinuity decreases its rate. Finally, our rate is lowered to only a few percent of
the computed value for the value of the free parameter $\beta$ employed, contrary to
almost all the implementations in the literature. This largely reduces spurious
pressure waves. More specifically, the equation describing the evolution of the
entropy includes a new term,

\begin{equation}
 \frac{\rm{d}A_i}{\rm{d}t}\stackrel{\rm{diff.}}{=}
 \frac{1}{\bar\rho^{\gamma-1}}\sum_j\alpha_{{\rm diff},ij} v_{{\rm
     diff},ij} \frac{m_j}{\rho_i + \rho_j} (\frac{\bar P_i}{\bar\rho_i} -
 \frac{\bar P_j}{\bar\rho_j}) \overline{W}_{ij},
\end{equation}
with the diffusion velocity given by $v_{{\rm diff},ij} = \max( c_i + c_j +
\left(\mathbf{v}_i-\mathbf{v}_j\right)\cdot\left(\mathbf{x}_i-\mathbf{x}
_j\right)/\left|\mathbf{x}_i-\mathbf{x}_j \right|, 0)$ and the diffusion coefficient
by $\alpha_{{\rm diff},ij} = \frac{1}{2}(\alpha_{{\rm diff},i} + \alpha_{{\rm
    diff},j})$. The individual diffusion coefficients are evolved alongside the other
thermodynamic variables following the differential equation

\begin{equation}
 \dot\alpha_{{\rm diff},i} = \beta \frac{h_i \nabla_i^2 \left(\bar
   P_i/(\gamma-1)\bar\rho_i\right)}{\sqrt{\bar P_i/(\gamma-1)\bar\rho_i}},
\end{equation}
where, as discussed above, we adopt $\beta=0.01$. We further impose $0<\alpha_{{\rm
    diff},i}<1$, but note that the upper limit is rarely reached, even for large
discontinuities.

\subsection{Thermal energy injection and time-step limiter}

A crucial aspect of the stellar feedback implementation used in \eagle and described
in \cite{DallaVecchia2012}, is the instantaneous injection of large amounts of thermal
energy $\Delta u$ in the ISM. This injection is performed by raising the temperature
of a gas particle by $\Delta T=10^{7.5}~\rm{K}$, a value much larger than the
average temperature of the warm ISM. In the \gadget formulation of SPH, this is
implemented by changing the entropy $A_i$ of a particle. In the case of \anarchy, the
situation is more complex since the densities themselves are weighted by the
entropies, which implies that a change in the entropy will affect both quantities
entering the equations of motion of all the particles in a given
neighbourhood. Hence, changing the internal entropy of just one single particle will
\emph{not} lead to the correct change of energy (across all particles in the
simulation volume) of the gas. The thermal energy injected in the gas will be
different (typically lower) from what is expected by a simple rise in $A_i$, leading
to a seemingly inefficient feedback event.

This problem is alleviated in the \eagle code by the use of a series of iterations
during which the values of $A_i$ and $\rho_i$ are changed until they have converged
to values for which the total energy injection is close to the imposed value:

\begin{align*}
 A_{i, n+1} &= \frac{(\gamma-1)(u_{\rm old} + \Delta u
   )}{\bar\rho_{i,n}^{\gamma-1}}, \\ \bar \rho_{i, n+1} &= \frac{\bar\rho_{i,n}
   A_{n}^{1/\gamma} -m_iW(0,h_i) A_{i,n}^{1/\gamma} + m_i W(0,h_i)
   A_{i,n+1}^{1/\gamma} }{A_{i, n+1}^{1/\gamma}}.
\end{align*}

This approximation is only valid for reasonable values of $\Delta u$ and only leads
to the injection of the correct amount of energy if the energy is injected into one
particle in a given neighbourhood, as is the case in most stellar feedback
events. This scheme typically leads to converged values (at better than the $5\%$
level) in one or two iterations. When large amounts of energy are injected into
multiple neighbouring particles, as can happen in some AGN feedback events, this
approximation is not sufficient to properly conserve energy (across all particles in
a given kernel neighbourhood). To avoid this, we limit the number of particles being
heated at the same time to $30\%$ of the AGN's neighbours. If this threshold is
exceeded, the time step of the BH is decreased and the remaining energy is kept for
injection at the next time step. Isolated explosion tests have shown that this limit
leads to the correct amount of energy being distributed.

As was pointed out by \cite{Saitoh2009}, the conservation of energy in SPH following
the injection of large amounts of energy requires the reduction of the integration
time-step of the particles receiving energy as well as those of its direct
neighbours. This was further refined by \cite{Durier2012}, who demonstrated that
energy conservation can only be achieved if the time-step of the particles is updated
according to their new hydrodynamical state. This latter time-step limiter is applied
in both the \gadget-SPH and \anarchy-SPH simulations used in sections
\ref{sec:galaxyPopulation} and \ref{sec:gasProperties} of this paper.  We discuss its
influence on galaxy properties in subsections \ref{ssec:gsmf} and \ref{ssec:sizes}.

\section{Galaxy population and evolution through cosmic time}
\label{sec:galaxyPopulation}

As discussed by \cite{Schaye2015} and \cite{Crain2015}, the subgrid models of stellar
and AGN feedback are only an incomplete representation of the physical processes
taking place in the unresolved multiphase ISM. In particular, because radiative
losses and momentum cancellation associated with feedback from star formation and AGN
in the multiphase ISM cannot be predicted from first principles, the
simulations cannot make ab initio predictions for the stellar and black hole masses.
In a fashion similar to the semi-analytic models, the subgrid models for feedback in
the \eagle simulations have therefore been calibrated to reproduce the $z=0.1$ galaxy
stellar mass function and the relations between galaxy size and mass and between the
mass of the central supermassive black hole and the galaxy. The details of this
calibration procedure are described in \cite{Crain2015}. In this section, we will
present the basic properties of our simulated galaxy population when the hydrodynamic
scheme is reverted to the commonly used \gadget-SPH formalism. We will specifically
focus on the galaxy stellar mass function and galaxy sizes before turning
towards the star formation rates.

We stress that the model parameters have not been recalibrated when switching 
our hydrodynamics scheme back to \gadget-SPH.

\subsection{The galaxy stellar mass function}
\label{ssec:gsmf}

\begin{figure}
\includegraphics[width=1.\columnwidth]{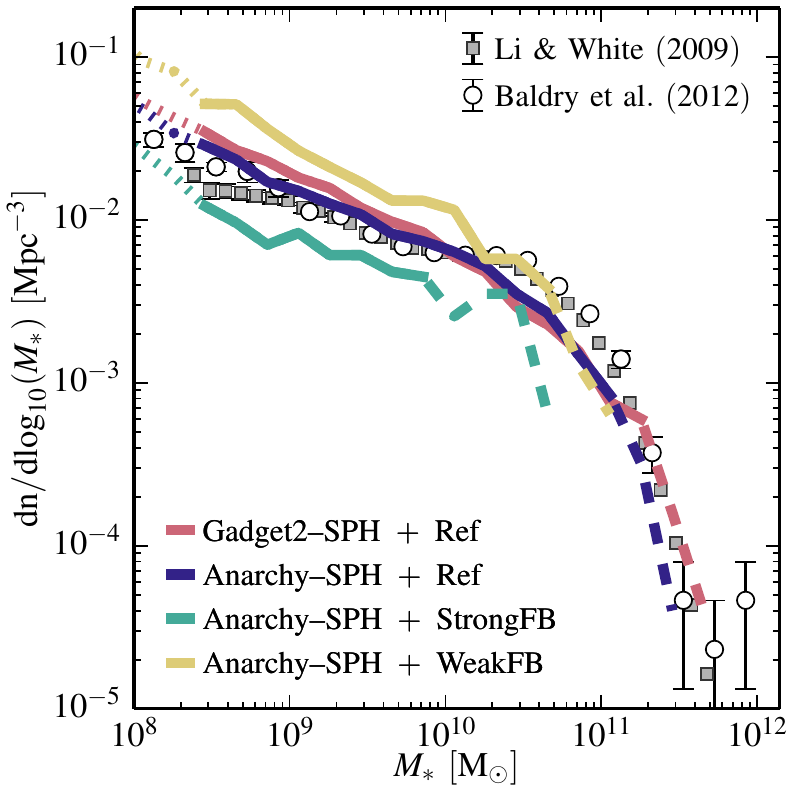}
\caption[The $z=0.1$ GSMF of the simulations using \anarchy SPH and \gadget SPH]{The
  $z=0.1$ GSMF of the L050N0752 simulations using \anarchy SPH (blue line, the \eagle
  default) and \gadget SPH (red line).  Curves are drawn with dotted lines where
  galaxies are comprised of fewer than 100 star particles, and dashed lines where the
  GSMF is sampled by fewer than 10 galaxies per $0.2~\rm{dex}$ mass bin. Data points
  show measurements with $1\sigma$ error bars from the SDSS \citep[][filled
    squares]{Li2009}, and GAMA \citep[][open circles]{Baldry2012} surveys. The yellow
  and green lines show the GSMF of the L025N0376 simulations with twice weaker and
  twice stronger feedback from star formation respectively, in a smaller
  $25^3~\rm{Mpc}^3$ volume.  The differences due to the choice of hydrodynamics
  scheme are smaller than the differences due to uncertainties in the sub-grid
  modelling.}
\label{fig:GSMF}
\end{figure}

In Fig. \ref{fig:GSMF}, we show the galaxy stellar mass function (GSMF) at $z=0.1$
computed in spherical apertures of $30~\rm{kpc}$ around the centre of potential of
the haloes. As discussed by \cite{Schaye2015}, this choice of aperture gives a simple
way to distinguish the galaxy and the ICL. The blue and red lines correspond to our
simulations with the \anarchy and \gadget flavours of SPH, respectively. We use
dashed lines when fewer than $10$ objects populate a ($0.2~\rm{dex}$) stellar mass
bin and dotted lines when the galaxy mass drops below our resolution limit \citep[for
  resolution considerations, see][]{Schaye2015}. The two hydrodynamic schemes lead to
very similar GSMFs with significant differences only appearing at
$M_*>2\times10^{11}~\msun$, where the small number of objects in the volume prevents
a strong interpretation of the deviation, based solely on that diagnostic. The white
circles and grey squares correspond to the observationally inferred GSMFs from the
GAMA \citep{Li2009} and SDSS \citep{Baldry2012} surveys, respectively. The two
simulated galaxy populations undershoot the break of the stellar mass function by a
similar amount and are in a similarly good agreement ($\lesssim0.2~\rm{dex}$) with
the data.  The choice of hydrodynamic solver seems to only impact the mass and
abundance of the most massive galaxies in our cosmological simulations.  We
re-iterate that there has been no recalibration of the subgrid parameters between the
\gadget and \anarchy simulations.

In order to compare the contribution of hydrodynamics uncertainties to the
uncertainties arising from the subgrid models, we show using green and yellow lines
two additional models using the \anarchy flavour of SPH but with feedback from star
formation injecting half and twice as much energy, respectively. These simulations
are the models WeakFB and StrongFB introduced by \cite{Crain2015} and reduced or
increased the number of feedback events taking place, whilst keeping the amount of
energy injected per event constant. They have been run in smaller volumes
($25^3~\rm{Mpc}^3$), leading to poorer statistics at the high-mass end. These changes
in the amount of energy injected in the ISM lead to much larger differences in the
GSMF than changing the flavour of SPH used for the simulation. 

The large impact of variations of the subgrid model for stellar feedback on the
simulated population and on single galaxies can also be appreciated from the large
range of outcomes of the different models in the \textsc{OWLS} suite
\citep{Schaye2010, Haas2013} and \textsc{Aquila} projects
\citep{Scannapieco2012}. Our work, however, uses a higher resolution than was
accessible in the \textsc{OWLS} suite for $z=0$ and contrary to \textsc{Aquila} uses
a cosmological volume and can hence study the effect of the hydrodynamics scheme from
dwarf galaxies to group-sized haloes.  The study of \cite{Keres2012}, which compared
the \textsc{arepo} \citep{Springel2010a} and \gadget-SPH hydro solvers but using
simple subgrid models, came to the same conclusion: the choice of hydrodynamics
scheme has little impact on the stellar mass function of simulated galaxies at
intermediate mass, only the most massive objects are affected. Interestingly, the
differences they observed in high mass galaxies are exactly opposite to our findings:
the more accurate solver (in their case \textsc{arepo}) produces more massive
galaxies than the simulation using \gadget-SPH. This confirms that the source terms
arising from the physical modelling of the unresolved processes in the ISM,
especially the modelling of AGN feedback (see the discussion below in Section
\ref{ssec:haloes}), clearly dominate the uncertainty budget.

We now turn to the impact of the time-step limiter on the simulated galaxy
population. As was demonstrated by \cite{Durier2012}, the absence of a time-step
limiter leads to the non-conservation of energy during feedback events. The energy of
the system after the injection is larger than expected. This implies that a
simulation without time-step limiter will have a spuriously high feedback
efficiency. In order to test this, we ran a simulation in a $25^3~\rm{Mpc}^3$ volume
using the Ref subgrid model and the \anarchy-SPH scheme but with the
\cite{Durier2012} time-step limiter switched off. Since this simulation volume is
too small to be representative, it is more informative to study the relation between
halo mass and stellar mass.

\begin{figure}
\includegraphics[width=1.\columnwidth]{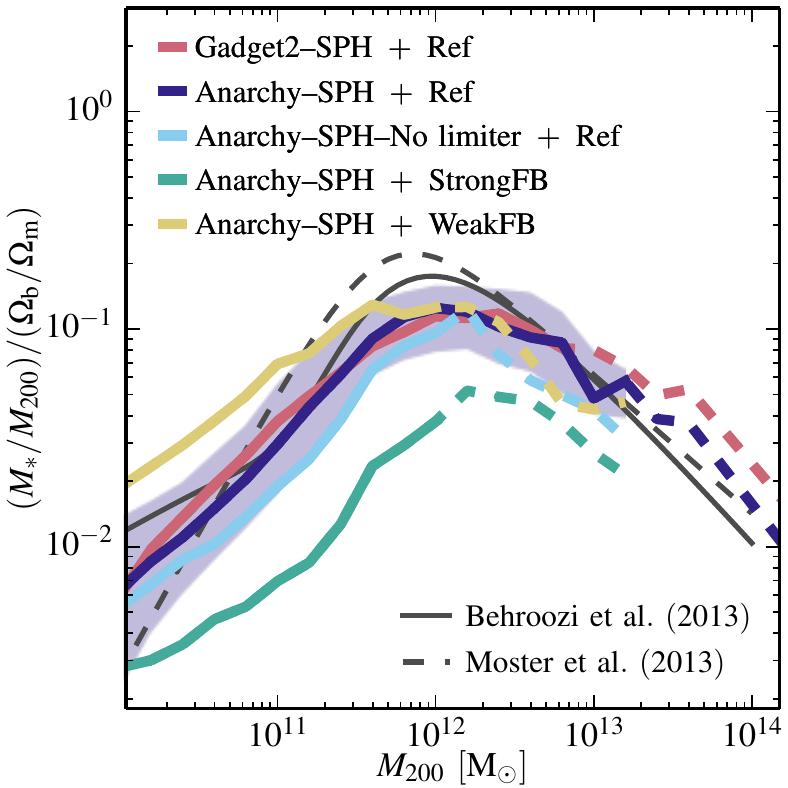}
\caption{The median ratio of the stellar and halo mass of central galaxies, as a
  function of halo mass $M_{200}$ and normalised by the cosmic baryon fraction at
  $z=0.1$ for both the L050N0752 \anarchy SPH (blue line) and \gadget SPH (red line)
  simulations.  Curves are drawn with dashed lines where the GSMF is sampled by fewer
  than 10 galaxies per bin. The $1\sigma$ scatter about the median of the \anarchy
  run is denoted by the blue shaded region. The solid and dashed grey lines show the
  multi-epoch abundance matching results of \citet{Behroozi2013} and
  \citet{Moster2013} respectively. The yellow and green lines show the GSMF of the
  L025N0376 simulations with twice weaker and twice stronger feedback from star
  formation respectively. The cyan line corresponds to the simulation using the
  \anarchy formulation of SPH and the reference subgrid model, but without the
  time-step limiter. The absence of the time-step limiter artificially increases the
  efficiency of the feedback and has a greater impact than the choice of hydro
  solver.}
\label{fig:no_limiter}
\end{figure}

In Fig. \ref{fig:no_limiter}, we therefore show the relation between halo mass
($M_{200}$) and galaxy formation efficiency ($M_*/M_{200}$) for central galaxies at
$z=0.1$. As for all other figures, the blue and red lines correspond to the
\anarchy-SPH and \gadget-SPH simulations respectively, both using the time-step
limiter. We show the simulation using twice stronger and twice weaker feedback with
green and yellow lines respectively. These are the same simulations that were shown
in Fig. \ref{fig:GSMF}. The stronger feedback from star formation leads to a lower
stellar mass formed in a given halo than in the Ref model, as was discussed by
\cite{Crain2015}. As expected from the GSMF, galaxy formation efficiency is strongly
moderated by the feedback parameters. Finally, we show in cyan the simulation using
the Ref subgrid model but without the time-step limiter. This simulation displays a
lower stellar mass in a given halo than its counterpart using the limiter. This
indicates that the feedback was indeed more efficient at quenching star formation in
that simulation, as expected from the analysis of \cite{Durier2012}. This is a purely
numerical effect that has to be corrected by the use of small time-steps in regions
where feedback takes place. The simulation volume considered for that test is too
small to contain a large sample of haloes hosting galaxies with significant AGN
activity. \cite{Durier2012} argued that the larger the energy jump, the larger the
violation of energy conservation will be when the time-step limiter is not used. As
the energy injection in AGN feedback events is two orders of magnitude larger than
for stellar feedback, we expect the masses of galaxies with
$M_*\gtrsim3\times10^{10}\msun$ (the mass range where AGN feedback starts to be
important) to be reduced, compared to the Ref model, even more than the galaxies for
which AGN feedback plays no role. Note that the impact of the time-step limiter is
much larger than the differences due to the hydrodynamics solver, but smaller than
the effect of doubling/halving the feedback strength.

\subsection{The sizes of galaxies}
\label{ssec:sizes}

\cite{Crain2015} showed that matching the observed GSMF does in general not lead
to a realistic population of galaxies in terms of their mass-size relation and mass
build up. Alongside galaxy masses, galaxy sizes were therefore considered in
the \eagle project during the calibration of the parameters of the subgrid model
for stellar feedback. \cite{Crain2015} demonstrated that numerical limitations
tend to make feedback from star formation less efficient at quenching the
galaxies if the feedback occurs in dense regions of the ISM. This would lead to
galaxies that are too compact and with a specific star formation rate at low
redshift that is lower than observed.  As a consequence, they also showed that
selecting model parameters that lead to galaxies with sizes in agreement with
observational data was necessary to obtain a realistic population of galaxies
across cosmic time. Assessing the dependence of the galaxy sizes on the
hydrodynamics scheme is, hence, crucial.

\begin{figure}
\includegraphics[width=1.\columnwidth]{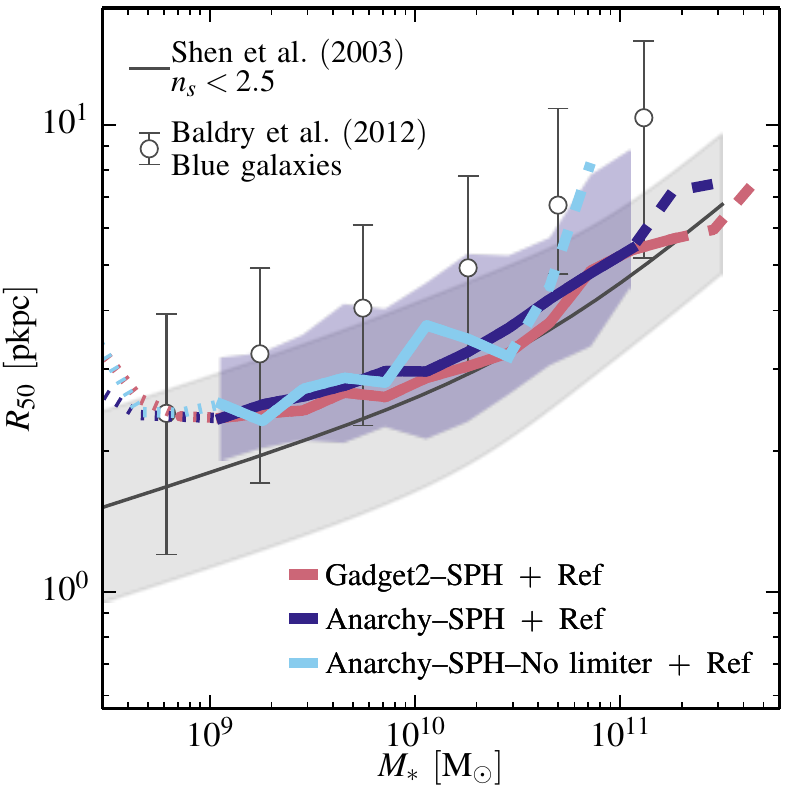}
\caption{The sizes, at $z = 0.1$, of disc galaxies in the L050N0752 \anarchy SPH
  (blue line) and \gadget SPH (red line) simulations and in the \anarchy SPH model
  without time-step limiter (cyan line). Size, $R_{50}$, is defined as the half-mass
  radius of a S\'ersic profile fit to the projected, azimuthally-averaged stellar
  surface density profile of a galaxy, and those with S\'ersic index $n_s < 2.5$ are
  considered disc galaxies. Curves show the binned median sizes, and are drawn with
  dotted lines below a mass scale of 600 star particles, and using a dashed line
  style where sampled by fewer than 10 galaxies per $0.2~\rm{dex}$ mass bin. The
  $1\sigma$ scatter about the median of the \anarchy run is denoted by the blue
  shaded region. The solid grey line and the grey shading show the median and
  $1\sigma$ scatter of sizes for $n_s < 2.5$ galaxies inferred from SDSS data by
  \citet{Shen2003}, whilst white circles with error bars show sizes of blue galaxies
  inferred by \citet{Baldry2012} from GAMA data. All simulations reproduce the
  $z=0.1$ galaxy sizes.}
\label{fig:Sizes}
\end{figure}

In Fig. \ref{fig:Sizes}, we show the sizes of the galaxies in both the \anarchy-SPH
and \gadget-SPH simulations. The observational data sets from \cite{Shen2003} (SDSS,
grey line and shading) and \cite{Baldry2012} (GAMA, white circles) are shown for
comparison. The sizes of the simulated galaxies are computed following
\cite{McCarthy2012}. We fit a S\'ersic profile to the projected, azimuthally-averaged
surface density profiles. We then extract the half-mass radius of the galaxy,
$R_{50}$, from this profile when integrated to infinity. To match the observational
selection of \cite{Shen2003}, we select only galaxies that have a S\'ersic index
$n_{\rm s}< 2.5$. We use dashed lines where the ($0.2~\rm{dex}$) mass bins contain
fewer than $10$ objects and dotted lines for galaxies that are represented by fewer
than $600$ star particles. The $1\sigma$ scatter around the mean in the \anarchy-SPH
simulation is shown as the blue shaded region for the mass bins that are both well
resolved and well sampled. The \gadget-SPH simulation presents a similar scatter.

Both simulations reproduce the observed galaxy size-mass relation. The simulated
galaxies lie within $0.1-0.2~\rm{dex}$ of either of the two data sets. As was the
case for the GSMF, the galaxy sizes are unaffected by the specific details of the
hydrodynamics scheme. This implies that the two hydro schemes have similar energy
losses in dense gas regions where feedback takes place. Differences much larger than
this can be seen when the subgrid model parameters are varied, even if one requires
the GSMF to match observations \citep{Crain2015}. Galaxies with $M_*>10^{11}\msun$
display small, but not statistically significant, differences with the objects in the
\gadget simulation being slightly smaller. This is in agreement with the findings of
\cite{Naab2007} who, using \gadget-SPH, produced massive galaxies too compact
compared to observations.

When considering the galaxy masses, we found that not using the \cite{Durier2012}
time-step limiter led to an increase of the feedback efficiency, although the
magnitude of the effect was small compared to that of doubling the feedback
energy. As galaxy sizes were our second diagnostic, we also consider the effect of
switching off this limiter on the sizes of our simulated galaxies. This model is
shown as the cyan line in Fig.~\ref{fig:Sizes}. The oscillations seen in the curve
are due to the smaller volume used for this simulation. The sizes of the galaxies are
very close to or slightly larger than the ones in the default simulation. 

\cite{Crain2015} also showed that using more efficient stellar feedback leads (among
other things) to higher SSFRs, lower passive fractions and lower metallicities. We
have verified that turning off the time-step limiter has the same qualitative
effects, although the differences are small.  We will not consider the effect of
turning off the limiter further in the rest of this paper.

\subsection{The star formation rates of galaxies}

We now turn to the star formation rates of galaxies. This quantity was not used in
the parameter calibration process of the \anarchy-SPH run (i.e. the default \eagle model)
and is an important independent diagnostic of the success of the
simulation. Furthermore, since the ISM dictates the star formation rates of galaxies,
changes in the way the equations of hydrodynamics are solved may lead to changes in
the SFRs.

\begin{figure}
\includegraphics[width=1.\columnwidth]{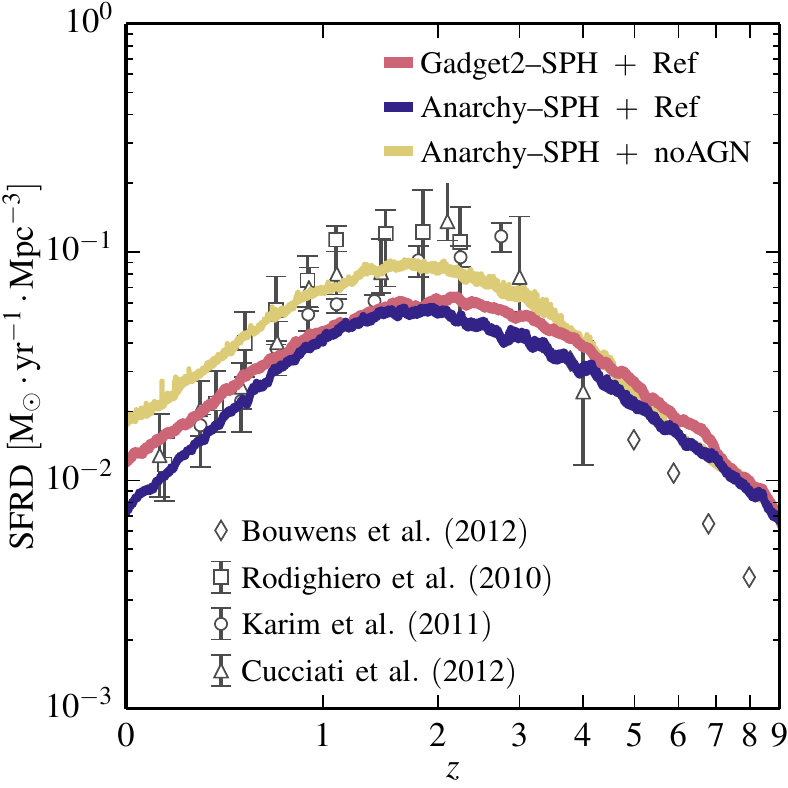}
\caption{The evolution of the cosmic star formation rate density in both the
  L050N0752 \anarchy SPH (blue line) and \gadget-2 SPH (red line) simulations. The
  data points correspond to observations from \citet{Karim2011} (radio),
  \citet{Rodighiero2010} ($24~\mu \rm{m}$), \citet{Cucciati2012} (FUV) and
  \citet{Bouwens2012} (UV). The decline in the star formation
  rate density from $z=2$ to $z=0$ is less pronounced in the \gadget run, leading to
  a $65\%$ higher star formation rate density at $z=0$. For comparison, a model
  without AGN feedback (yellow line) is shown. The star formation rate density in
  that model has a low-redshift slope similar to that of the \gadget simulation.}
\label{fig:Madau}
\end{figure}

In Fig. \ref{fig:Madau}, we show the average star formation rate per unit volume. The
blue and red lines again correspond to the \anarchy and \gadget flavours of SPH,
respectively. Observational data from \cite{Rodighiero2010}, \cite{Karim2011},
\cite{Cucciati2012} and \cite{Bouwens2012} is also shown. Where applicable, the data
has been corrected for our adopted cosmology and IMF as described in
\cite{Furlong2015}. In agreement with the data, both simulations display a rise in
the star formation rate density at high redshifts and a fall at $z\lesssim2$. As was
discussed by \cite{Furlong2015}, the constant offset in star formation rate of
$\approx0.2~\rm{dex}$ between the simulations and observations leads to $20\%$ less
stars being formed over the cosmic history, consistent with the $z=0.1$ GSMF
(Fig. \ref{fig:GSMF}), whose ``knee'' the simulations slightly undershoots.

The simulation using the \gadget version of SPH predicts a higher cosmic star
formation rate density than its \anarchy counterpart between redshifts $2$ and $6$
but this does not lead to a large difference in stellar mass formed by
$z=2$. However, the higher star formation rate seen at $z<1$ is important and the
smaller decrease between $z=1$ and $z=0$ implies a star formation rate that is $65\%$
higher by $z=0$ in the simulation using the \gadget formulation of SPH. This higher
star formation rate can be tentatively related to the larger number of high-mass
galaxies seen in the GSMF of this simulation and could, hence, indicate a lower
quenching efficiency of the AGN activity in the largest haloes. An extreme version of
a model with a low quenching efficiency in large haloes is given by a model without
AGN feedback.  Such a model, using the \anarchy flavour of SPH, is shown using the
yellow line in Fig.~\ref{fig:Madau}. The excess star formation at $z<2$ is much
larger than in the \gadget-SPH based run with AGN feedback, but the slope is similar
and not steep enough compared to the data.

\begin{figure}
\includegraphics[width=1.\columnwidth]{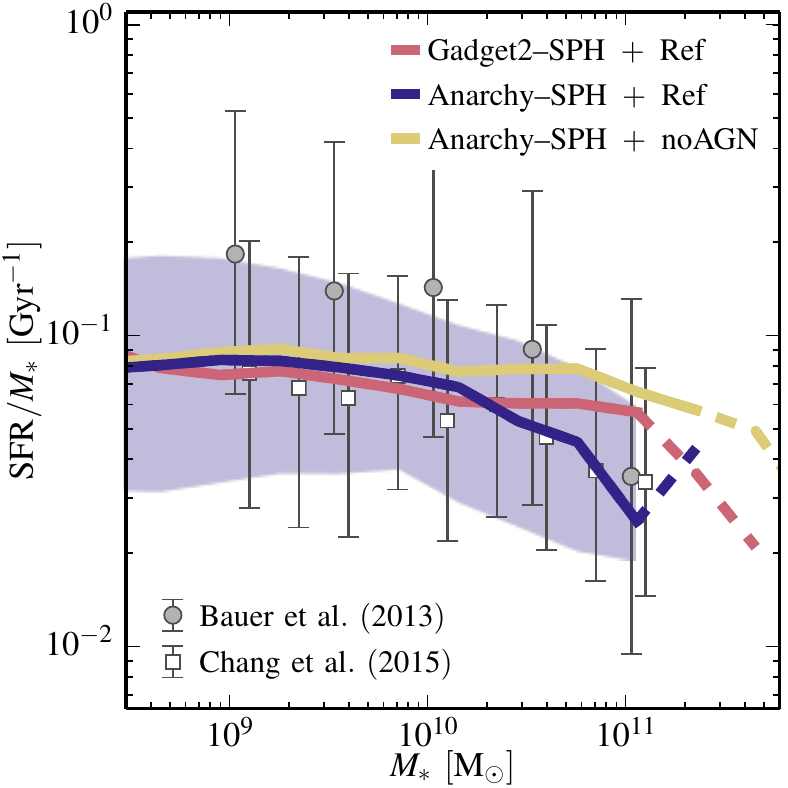}
\caption{The median specific star formation rate $\dot M_*/M_*$, of star-forming
  galaxies ($\dot M_*/M_* > 0.01~\rm{Gyr}^{-1}$) as a function of stellar mass at
  $z=0.1$ in the L050N0752 \anarchy SPH (blue line) and \gadget-2 SPH (red line)
  simulations. Dashed line styles are used where the simulation is sampled by fewer
  than 10 galaxies per $0.2~\rm{dex}$ mass bin. The $1\sigma$ scatter about the
  median of the \anarchy run is denoted by the blue shaded region.  Observational
  data points with error bars correspond to the median and $1\sigma$ scatter of the
  SSFR from GAMA by \citep[][grey circles]{Bauer2013} and SDSS+WISE by \citep[][white
    squares]{Chang2015}. Galaxies with $M_*>2\times10^{10}\msun$ have a significantly
  higher specific star formation rate in the \gadget SPH simulation than in the
  \anarchy SPH one, but the decrease is smaller than when AGN activity is turned off
  (yellow line).}
\label{fig:SSFR}
\end{figure}

Whether the excess star formation rate at low redshift is due to large haloes can be
confirmed by looking at the specific star formation rate (SSFR) of the simulated
galaxies.  This quantity is shown in Fig. \ref{fig:SSFR} as a function of stellar
mass. We limit our selection to star-forming galaxies by excluding objects with $\dot
M_*/M_* < 0.01~\rm{Gyr}^{-1}$. As was the case for the stellar mass of the galaxies,
we measure the SFR within a $30~\rm{kpc}$ spherical aperture. The red and blue lines
show the mean SSFR in the simulations using the \gadget and \anarchy flavours of SPH
respectively. As for other figures, the lines are dashed when a given mass bin is
sampled by fewer than $10$ objects. The blue shaded region indicates the
$1\textendash\sigma$ scatter in the \anarchy-based simulation. The \gadget-based
simulation displays a scatter of the same magnitude. For comparison, we show the SSFR
inferred from observations in the GAMA survey by \cite{Bauer2013} (grey circles) and
observations by \cite{Chang2015} using recalibrated star formation rate indicators
based on SDSS+WISE photometry (white squares). Simulated galaxies with masses
$M_*\sim10^{11}\msun$ are in agreement with the \cite{Bauer2013} data, whilst
lower-mass objects exhibit a specific star formation rate lower than observed with
the discrepancy reaching $\sim0.3~\rm{dex}$ at $M_*\sim10^9\msun$. \cite{Schaye2015}
showed that part of this discrepancy goes away if the resolution of the simulation is
increased. Interestingly, the recalibrated star formation tracers of \cite{Chang2015}
lead to lower specific star formation rates in excellent agreement with the \eagle
results. Both the \gadget and \anarchy simulations show the same behaviour at low
masses.

At the upper end of the mass spectrum the two simulations do, however, differ. The
star formation rate of galaxies with $M_*\gtrsim2\times10^{10}\msun$ is significantly
larger for the \gadget formulation of SPH. At $M_*\sim10^{11}\msun$, the discrepancy
is $0.3~\rm{dex}$.

\begin{figure}
\includegraphics[width=1.\columnwidth]{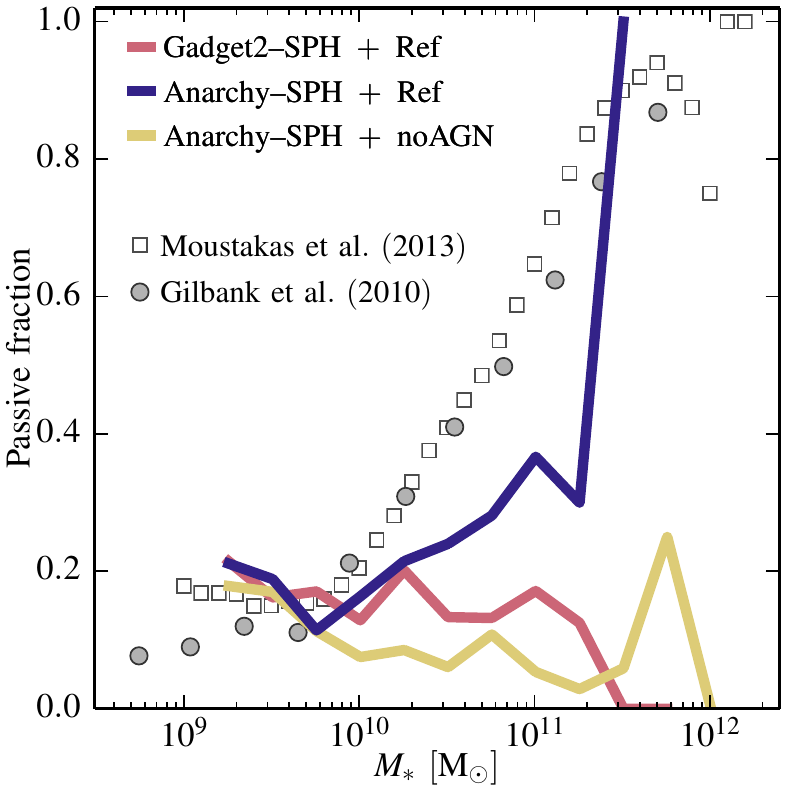}
\caption{The fraction of passive galaxies ($\dot M_*/M_* < 0.01~\rm{Gyr}^{-1}$) at
  $z=0.1$ in both the L050N0752 \anarchy SPH (blue line) and \gadget SPH (red line)
  simulations. We only show mass bins that correspond to 100 or more star-forming
  particles for the median SSFR. The grey circles and white squares correspond to the
  passive fractions inferred from the SDSS data by \citet{Gilbank2010} and by
  \citet{Moustakas2013}. The passive fraction is far too low for galaxies with
  $M_*\gtrsim2\times10^{10}\msun$ in the \gadget simulation, in a similar fashion to
  the \anarchy simulation without AGN feedback (yellow line).}
\label{fig:passive}
\end{figure}

Complementary to the SSFR of the star forming galaxies, the passive fraction provides
a good diagnostic of the efficiency with which SF is quenched in large galaxies This
quantity is shown in Fig. \ref{fig:passive} for both our simulations. Galaxies are
considered passive if their SSFR is smaller than $0.01~\rm{Gyr}^{-1}$, which is an
order of magnitude below the observed median SSFR for star forming galaxies at that
redshift. For comparison, the data points show the fractions inferred from SDSS data
by \cite{Gilbank2010} and \cite{Moustakas2013}. We only show points for the simulated
population at masses for which there are at least $100$ particles at the median SSFR
\citep[see][]{Schaye2015}.

The two simulations present a very different behaviour for galaxies with
$M_*\gtrsim2\times10^{10}\msun$. Whilst the \anarchy-SPH simulation follows the trend
seen in the observational data, the \gadget-SPH simulation shows a constant passive
fraction of $\sim15\%$ at masses up to $M_*=2\times10^{11}\msun$. At larger masses,
the fraction is $0$, implying that all galaxies are star-forming, in disagreement
with the data that indicates that almost all galaxies ($>80\%$) of that mass range
are passive. Note, however, that there are only $20$ galaxies with $M_*>10^{11}\msun$
in the simulation volume and that the fractions displayed in Fig. \ref{fig:passive}
are, hence, affected by small number statistics. Since the \anarchy and \gadget
simulations use the same initial conditions, the comparison between the two schemes
is, however, still meaningful. Switching from \anarchy to standard \gadget has
qualitatively a similar effect as switching off AGN feedback (yellow line).

The shortage of passive galaxies in the \gadget simulation at the high-mass end of
the galaxy population and the higher SSFR for high-mass objects both indicate that
the star formation quenching processes are inefficient in the largest haloes. This
higher star formation rate at low redshift in high-mass haloes leads to an increase
of the stellar mass of massive galaxies as was hinted at by the difference in the
GSMFs between the two simulations at $z=0.1$ (Fig.  \ref{fig:GSMF}). AGN feedback,
which is the main source of quenching in our model for galaxies with
$M_*\gtrsim2\cdot10^{10}\msun$, seems to be insufficiently effective at quenching
star formation in large haloes in the \gadget simulation.

It is worth mentioning that we cannot eliminate the possibility that a re-calibration
of the subgrid parameters could bring the \gadget simulation into agreement with the
data. By changing the frequency of the AGN events or the temperature to which the gas
is heated during such an event, it might be possible to quench star formation in
large galaxies even when the \gadget formulation of SPH is used. It is, however,
unclear if this could be achieved and whether subgrid parameters should be used to
compensate for the shortcomings of a particular hydro scheme. Similarly, simulations
run at different resolutions might lead to different conclusions (if the subgrid
parameters are kept fixed). Note that simulations run at a lower resolution (such as
the low-redshift versions of \textsc{OWLS} \cite{Schaye2010} and cosmo-\textsc{OWLS}
\citep{LeBrun2014}) have fewer resolution elements in the haloes and may hence not
suffer as much from the lack of phase mixing (see discussion below). A full
exploration of the subgrid model parameter space or a comprehensive resolution study
are, however, beyond the scope of the present paper.

The effectiveness of the AGN feedback can be related to the state of the gas
surrounding the galaxies and in the whole halo. The difference can be understood as
follows. The accretion of cold gas onto the galaxies from filaments is the key source
of fresh material from which stars can be formed in those haloes. The AGN will
sustain a hot halo in which these filaments will dissolve. It is likely that the
spurious surface tension that plagues the density-entropy formulation of SPH used in
\gadget does not leave the gas in the hot halo in a state where the AGN activity can
be effective at stopping star formation.  An example of these issues would be the
inability for dense gas blobs to dissolve in a hot halo medium \citep[see for
  instance the ``blob test'' problem by][]{Agertz2007}, which could allow cold
pristine gas in filaments to survive the hot bubbles created by the AGN activity and
feed the galaxy with gas ready to form stars. The better phase-mixing ability of the
\anarchy formulation of SPH is more effective at disrupting infalling filaments and
prevent them from reaching the galaxies, making the AGN-driven bubbles effective at
stopping star formation. In this scenario, the issue is not that outflows generated
by an AGN are unable to sustain a hot halo (we will show that hot haloes are present
in both cases), it is rather the pristine gas that forms clumps that are unstable and
cool rather than being mixed in. The next section further explores the differences in
gas properties of the two simulations.

\section{Large- and small-scale gas distribution}
\label{sec:gasProperties}

\begin{figure*}
\includegraphics[width=1.\columnwidth]{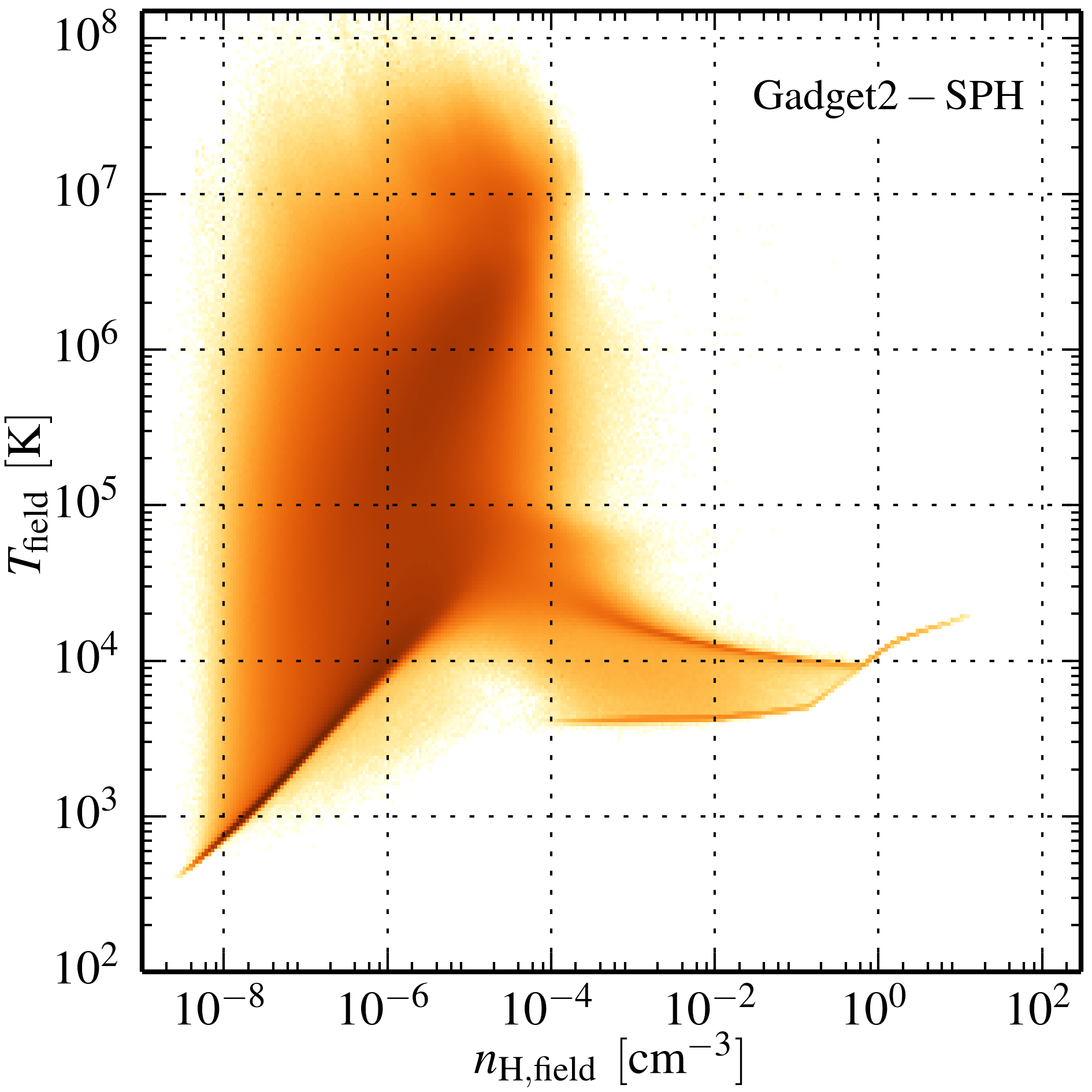}
\includegraphics[width=1.\columnwidth]{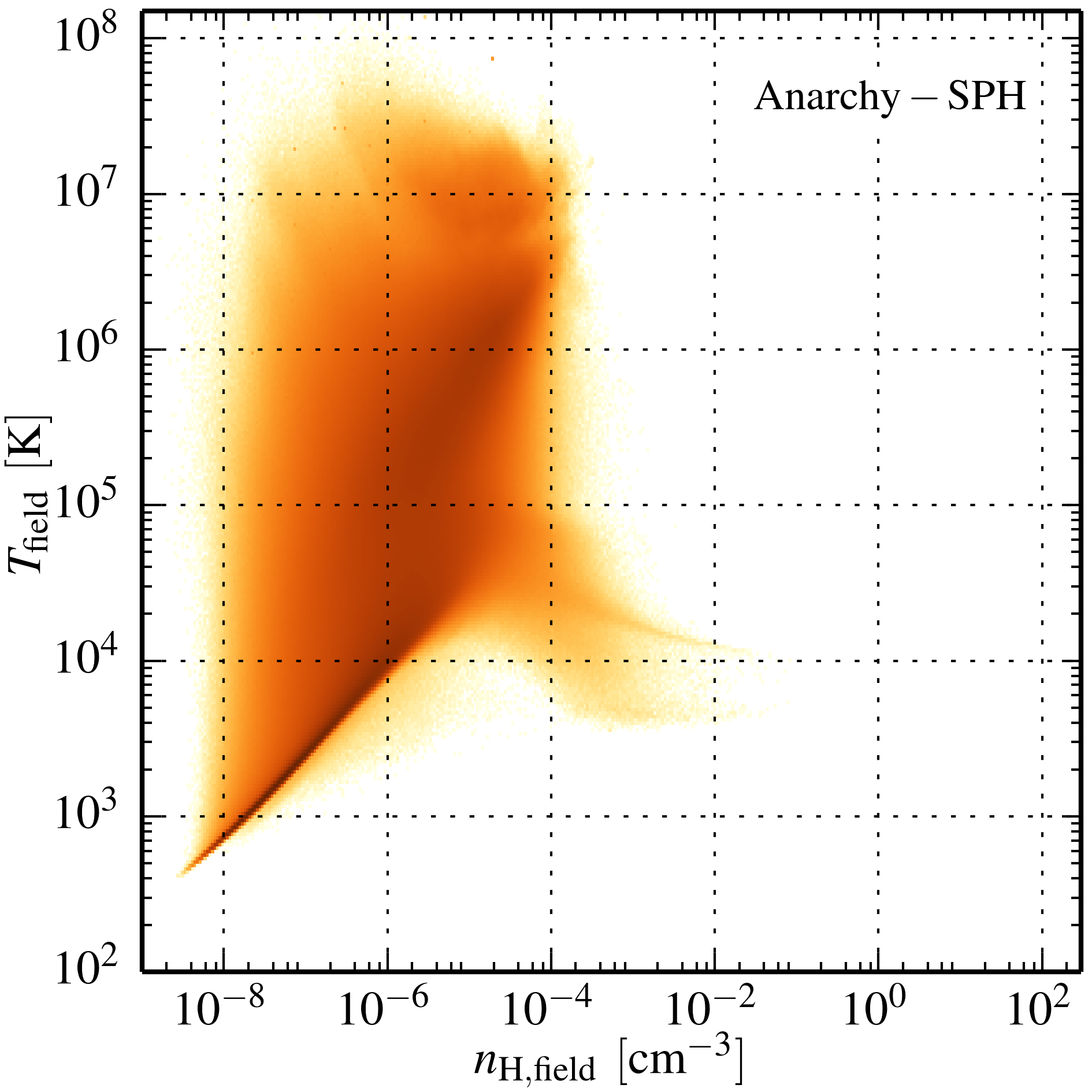}
\caption{The mass-weighted distribution of gas outside of collapsed structures in the
  density-temperature plane. The left panel shows the $z=0$ distribution for the
  \gadget SPH simulation whereas the right panel shows the equivalent distribution
  for the \anarchy SPH simulation. The \gadget SPH run displays high-density gas on
  the imposed equation of state whilst there is no gas in the \anarchy SPH run above
  a density of $n_{\rm H}> 10^{-1} \rm{cm}^{-3}$. Dense star forming gas is mixing
  with the lower-density, higher-temperature medium in the \anarchy SPH run, whilst
  the artificial surface tension introduced by the \gadget SPH formulation prevents
  this gas from dissolving and leads to star formation outside of haloes. }
\label{fig:rhoT_LSS}
\end{figure*}

In the previous section, we showed that the masses and sizes of galaxies are only
marginally affected by the improvements to the hydrodynamics scheme made in the
\anarchy flavour of SPH. We also showed, however, that the star formation rates of
massive galaxies are significantly affected by these same improvements and argued that
some of the differences might be directly related to the way in which the different
SPH schemes treat the gas in large haloes. In this section, we explore this
possibility by studying the state of the gas both outside and inside haloes. We will
focus on the largest systems, where the dynamical time is similar to or shorter than the
cooling time of the hot gas, and hence the hydrodynamic forces become important.

\subsection{Gas in large-scale structures}
\label{ssec:filaments}

A simple diagnostic of the state of the gas in a simulation is the distribution of
the SPH particles or grid cells in the density-temperature plane. The different
components (ISM, IGM, etc.) can then be identified and their relative abundance in
terms of mass or volume estimated. Since the \anarchy and \gadget formulations behave
differently when different phases are in contact or in the presence of a shock, it is
worth analysing the differences created by those schemes. In order to minimize the
impact of the subgrid models on the distribution of the gas, we start by looking at
the gas in the inter-halo medium, i.e. the gas outside of haloes.  Most of the gas
that is located outside of haloes has had little contact with star forming regions or
with the winds driven by AGN and star formation but some of the material might have
been enriched early on in proto-haloes \citep[e.g.][]{Oppenheimer2012}. We are hence
focusing on the low-metallicity, mostly primordial, gas before it falls onto
haloes. This should allow us to consider differences driven mostly by the two
flavours of the hydrodynamics scheme.

The haloes have been identified using the FoF algorithm and are hence typically
larger than the commonly given virial radii. This ensures that we are not considering
particles that are part of any resolved haloes. In both our simulations, we only
identify haloes that have more than $32$ particles, effectively imposing a minimum
halo mass of $M_{\rm FoF}=3.1\times10^8\msun$. This analysis is resolution dependent
via the definition of the minimum halo mass resolved by the simulation. If the
resolution were increased, one would find smaller haloes, meaning that some of the
particles that we identify as being outside of any halo will become part of small
haloes.  However, small haloes are unlikely to host large amounts of star formation
and drive enrichment and feedback. As both simulations have been run at the same
resolution with the same initial conditions, the same objects will collapse and form
haloes, ensuring that our one-to-one comparison is not compromised by the potential
presence of smaller unresolved structures.

In Fig. \ref{fig:rhoT_LSS}, we show the distribution of the gas outside of all FoF
groups in the density-temperature plane at $z=0$ for \gadget-SPH (left panel) and
\anarchy-SPH (right panel). The low-density material ($n_{\rm
  H}<10^{-4}~\rm{cm}^{-3}$) is in a very similar state in the two simulations with an
extended distribution of diffuse material spanning more than $4$ orders in magnitude
in temperature. The higher temperature material has been heated by feedback activity
and blown out of the haloes in both simulations. Differences start to appear at
intermediate densities ($10^{-4}~\rm{cm}^{-3}<n_{\rm H}<10^{-1}~\rm{cm}^{-3}$). A lot
more mass resides in that regime in the simulation using the \gadget formulation of
SPH. Because of the artificial surface tension appearing in \gadget-SPH between
different phases in contact discontinuities, this dense gas is unable to properly mix
with the lower density, higher temperature material surrounding it. In the \anarchy
simulation, the use of both the Pressure-Entropy formulation of the SPH equations and
of a (small numerical) diffusion term has allowed this dense gas to dissolve into its
surroundings. The difference is even more striking at higher densities ($n_{\rm
  H}>10^{-1}~\rm{cm}^{-3}$), where no gas is present in the \anarchy simulation,
whilst a significant amount is present in the \gadget one. This difference is
especially important since, depending on its metallicity, some of this dense gas may
be star-forming. Star formation is hence taking place outside of collapsed structures
in the simulation using \gadget. Interestingly, this high-density gas also has a high
metallicity ($Z\gtrsim0.1Z_\odot$). This gas has thus been ejected from haloes after
having been enriched by star formation. In \anarchy-SPH, similar material would
likely be dissolved into the surrounding lower-density medium, either outside haloes
or in winds inside haloes.

\subsection{Extragalactic gas in haloes}
\label{ssec:haloes}

We find that within haloes differences in the density-temperature diagram are best
quantified by looking at the distribution of star forming gas.  We define the
IntraGroup Medium (IGrM) as the gas within $R_{200}$ but outside of $30~\rm{kpc}$
masks placed at the centre of each subhalo. This excludes the gas present in the ISM
or close to galaxies and should leave us with a reasonable definition of the IGrM.

\begin{figure}
\includegraphics[width=1.\columnwidth]{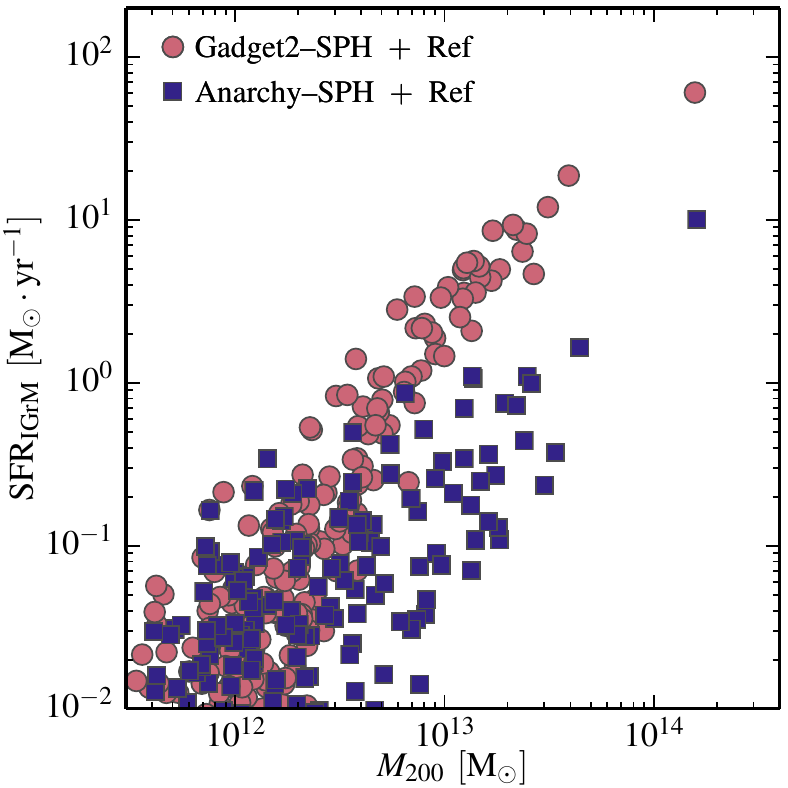}
\caption{The star formation rate of the inter-group medium (IGrM), i.e. inside the
  halo but at least $30~\rm{kpc}$ from any identified galaxy, as a function of halo
  mass at $z=0.1$ for the L050N0752 \anarchy SPH (blue squares) and \gadget SPH (red
  circles) simulations.  The IGrM is forming significantly more stars in group- and
  cluster-mass haloes ($M_{200}>5\times10^{12}\msun$) in the run using the \gadget
  SPH scheme.}
\label{fig:M200_SFR}
\end{figure}

In Fig. \ref{fig:M200_SFR}, we show the star formation rate of the IGrM as a function
of the halo mass $M_{200}$ at $z=0.1$ for objects extracted from the \anarchy
simulation (blue squares) and the \gadget-SPH simulation (red circles). Haloes with
masses $M_{200}\gg10^{12}\msun$ have a higher star formation rate in the IGrM in the
simulation using the \gadget formulation of SPH than in the \anarchy simulation. The
higher fraction of dense gas ($n_{\rm H}>10^{-1}~\rm{cm}^{-3}$) in the \gadget
simulation leads to a higher IGrM star formation rate. The specific star formation of
the IGrM corresponds to $\approx5\times10^{-3}~\rm{Gyr}^{-1}$ in the \gadget
simulation and is more than an order of magnitude lower
($\approx4\times10^{-4}~\rm{Gyr}^{-1}$) for \anarchy. Although these values are low
when compared to the typical values for galaxies (see Fig. \ref{fig:SSFR}), the
presence of significant star formation in the IGrM indicates that the AGN activity or
gravitational heating is not effective enough at quenching star formation in the
largest haloes.

As the haloes in the \gadget-based simulation exhibit more star formation in their
IGrM, it is interesting to investigate how the dense gas is distributed spatially. To
this end, we selected the most massive halo ($M_{200}\approx2\times10^{14}\msun$) in
both simulations and constructed column density maps of the gas. As we are mainly
interested in the dense gas and to increase the clarity of the maps, we only select
gas with $n_{\rm H} > 10^{-2}~\rm{cm}^{-3}$. As discussed above, the behaviour of the
warm diffuse medium is similar for both formulations of the SPH equations and can
hence be safely discarded here.

\begin{figure*}

\includegraphics[width=1.\columnwidth]{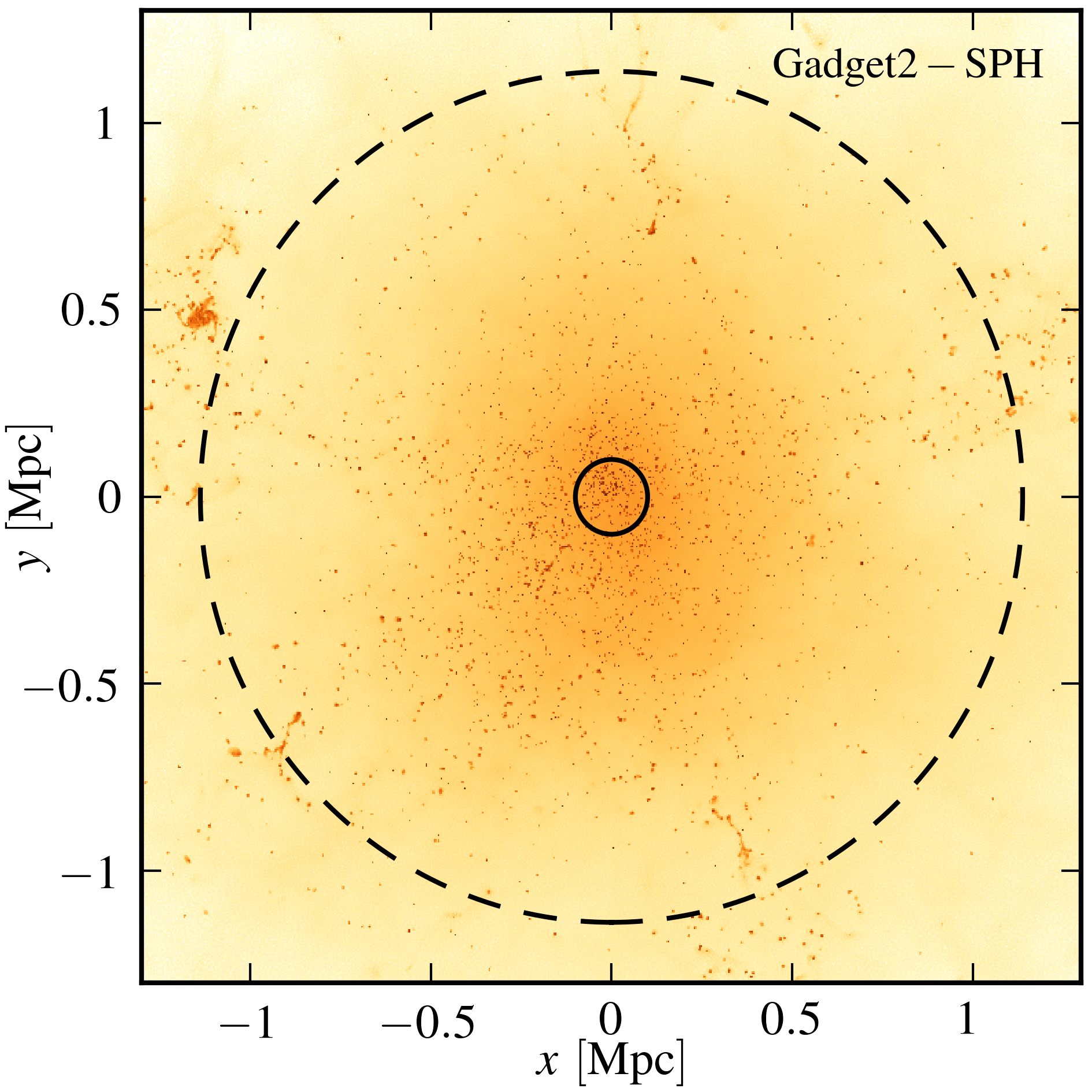}
\includegraphics[width=1.\columnwidth]{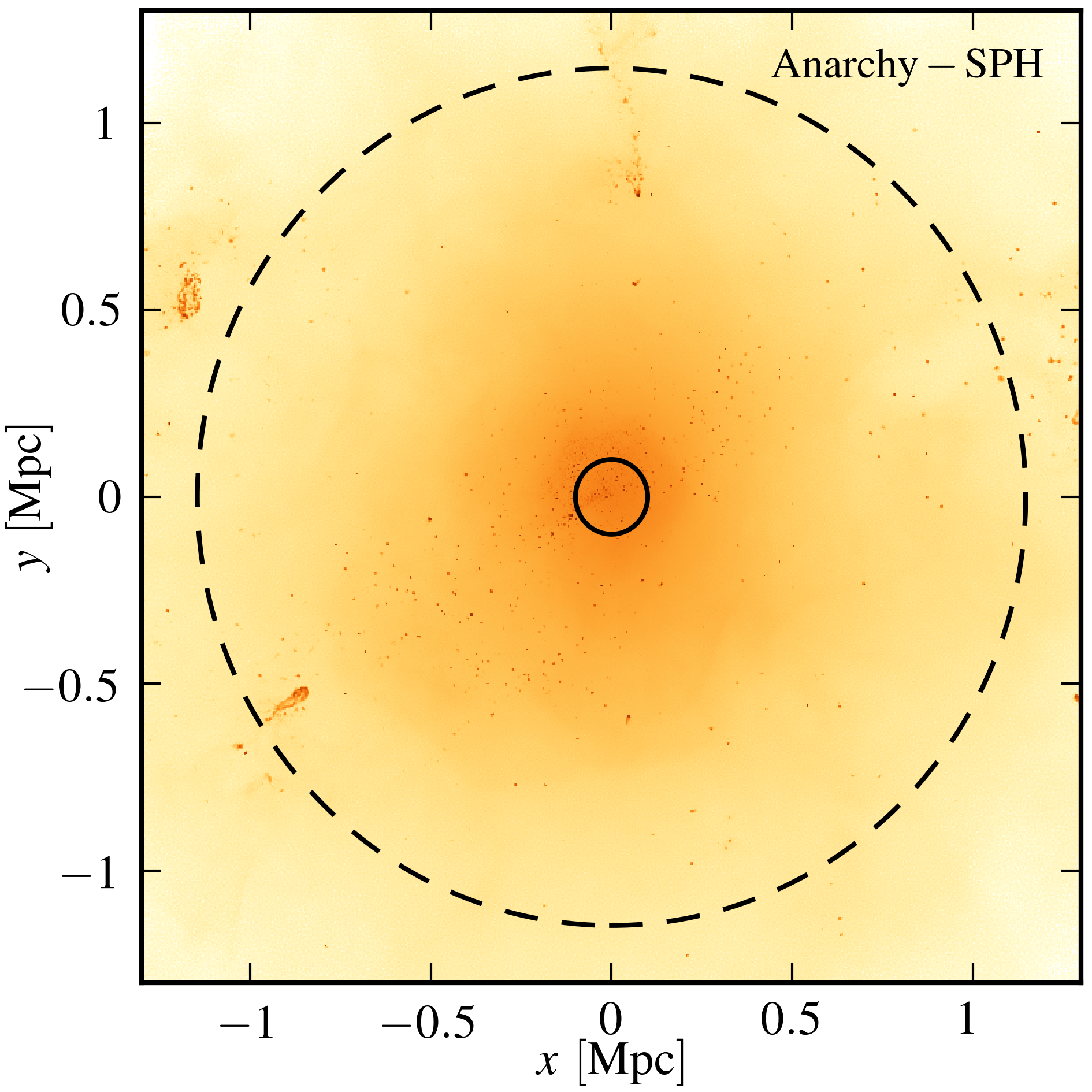}
\caption{Maps of the column density of dense gas ($n_{\rm H} > 0.01~\rm{cm}^{-3})$ in
  the largest haloes ($M_{200} \approx2\times10^{14}\msun$) of the L050N0752 \gadget
  SPH (left panel) and \anarchy SPH (right panel) simulations. The large dashed
  circle shows the location of the spherical overdensity radius $R_{200}$, whilst the small
  solid circle in the centre encloses the inner $100~\rm{kpc}$. The halo in the
  \gadget SPH run contains a large number of dense clumps of gas, as was found
  by \citet{Kaufmann2009} in their simulations, while its counterpart in the \anarchy
  SPH run displays a much smoother gas distribution. The spurious surface tension
  appearing in the \gadget formulation of SPH makes it difficult for the dense gas
  stripped from the in-falling satellites to be disrupted and mixed into the IGrM.}
\label{fig:maps}
\end{figure*}

These dense gas column density maps are shown in Fig. \ref{fig:maps} for the \gadget
(left panel) and \anarchy (right panel) simulations. The large dashed circles
indicate the position of the spherical overdensity radius,
$R_{200}\approx1.1~\rm{Mpc}$, whilst the small solid circles indicate the innermost
$100~\rm{kpc}$, where the effects of the central galaxies on the gas will be
maximized. We will not consider this central region in the remainder of this
subsection since, as was discussed in section \ref{sec:galaxyPopulation}, in this
region the differences due to the hydro solver are likely to be smaller than the ones
induced by small variations in the subgrid parameters.

The difference between the two maps is striking. The halo from the \gadget simulation
contains a large number of dense clumps of gas at all radii, as was found in the
simulations of \cite{Kaufmann2009}. These clumps can be seen even inside the inner
$100~\rm{kpc}$ where feedback from both the AGN and star formation might be expected
to disrupt them. These nuggets of dense gas also accompany the in-falling
satellites. The map extracted from the \anarchy simulation is much smoother and dense
gas is found mostly in the wakes of in-falling satellite galaxies following their
stripping.  \anarchy's ability to mix phases in contact discontinuity allows dense
clumps to dissolve into the hot halo, whereas the spurious surface tension that
appears between phases in \gadget SPH allows them so survive and perhaps even
grow. Since some clumps reach densities that exceed the threshold for star formation,
some of them will increase the SF rate of the IGrM. Here, the flavour of SPH has a
direct consequence on the observables extracted from the simulation.

\begin{figure}
\includegraphics[width=1.\columnwidth]{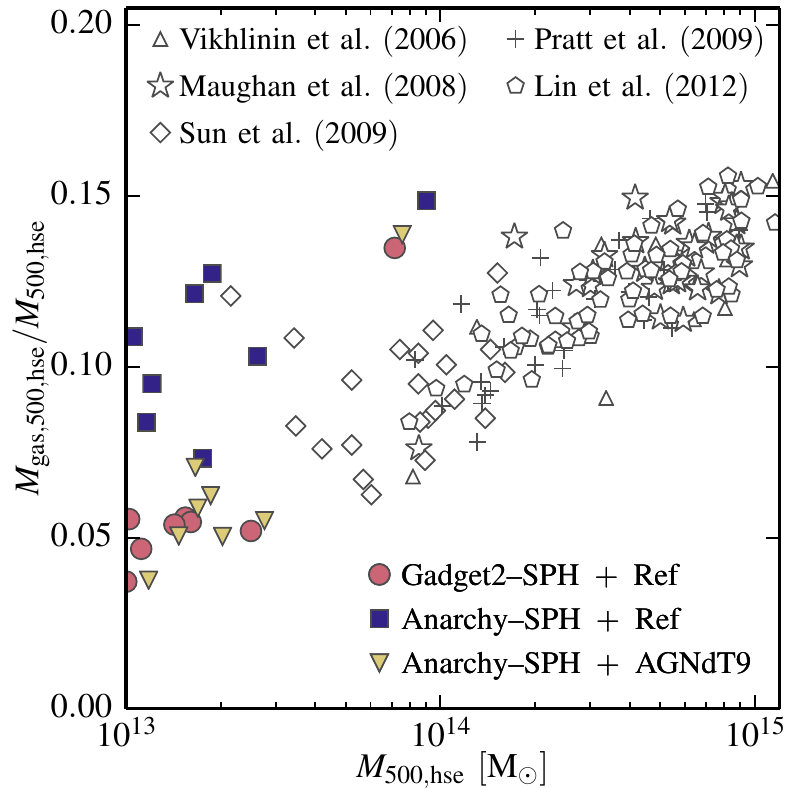}
\caption{The $z = 0$ gas fractions within $R_{500,\rm{hse}}$ as a function of
  $M_{500,\rm{hse}}$ inferred from virtual X-ray observations of the L050N0752
  \anarchy SPH (blue squares) and \gadget SPH (red circles) simulations.  Data points
  correspond to measurements from \citet{Vikhlinin2006} (triangles),
  \citet{Maughan2008} (stars), \citet{Sun2009} (diamonds), \citet{Pratt2009}
  (crosses) and \citet{Lin2012} (pentagons). The \anarchy SPH Ref model overpredicts
  the gas fractions for group-sized objects but this can be solved by using the
  AGNdT9 prescription for AGN feedback (yellow triangles). The haloes of the \gadget
  SPH run are in better agreement with the data as a result of their higher fraction
  of cold gas that artificially reduces the X-ray inferred gas fractions.}
\label{fig:gas_M500}
\end{figure}

Another observable that may be affected by the choice of hydrodynamics scheme is the
gas fraction. In Fig. \ref{fig:gas_M500}, we show the result of mock X-ray
observations of our haloes. Following the method described in \cite{LeBrun2014}, we
realise mock X-ray observations of our haloes and, assuming hydrostatic equilibrium,
infer the halo mass and gas fraction following the same analysis that is applied to
observational data. For comparison, we show data from \cite{Vikhlinin2006},
\cite{Maughan2008}, \cite{Sun2009}, \cite{Pratt2009} and \cite{Lin2012}. We only
selected clusters at $z<0.25$. As was discussed by \cite{Schaye2015}, the simulation
using the \anarchy flavour of SPH (blue squares), the Ref model of \eagle, overshoots
the extrapolated trend seen for higher-mass haloes. This indicates either that the
amount of X-ray gas in these haloes is too high or that the gas is in the wrong
thermodynamic state. The analysis of a larger simulation volume with more haloes
overlapping with the observations motivated \cite{Schaye2015} to introduce an
alternative model (labelled AGNdT9) for which the mock-observation inferred gas
fractions are in better agreement with the trend in the data. This model uses more
sparse, but also more energetic AGN heating events and is shown in
Fig. \ref{fig:gas_M500} using yellow triangles\footnote{We note that the map of the
  column density of dense gas of the largest halo in this model is very similar to
  the one using the Ref model and the \anarchy code (Fig.~\ref{fig:maps}, right
  panel). There is no large pool of dense gas clumps floating in the halo.}.

Interestingly, the \eagle Ref model using the \gadget version of SPH (red circles)
yields results that are very similar to the improved AGNdT9 model combined with
\anarchy-SPH. The gas fractions are in reasonable agreement with the data. However,
the analysis of the dense gas maps and the following discussion indicates that this
better agreement is mostly accidental and not a success of the model. The X-ray
inferred gas fractions are driven down by a change in the gas mass in the haloes but
also by the presence of cold and dense gas in the IGrM that does not emit X-ray and
hence artificially reduces the inferred gas masses. The cold clumps lead to the star
formation seen in Fig.~\ref{fig:rhoT_LSS}. We note, however, that these spurious
undisrupted clumps of dense gas are unlikely to affect simulations of the IGrM done
at lower resolution such as those of \cite{McCarthy2010} or
\cite{LeBrun2014}. Spurious surface tension, preventing the mixing of phase, only
appears when $\mathcal{O}(10)$ particles are part of a cold gas fragment. In
lower-resolution simulation such gas blobs are sampled by fewer particles and will
mix with their environment.

\begin{figure}
\includegraphics[width=1.\columnwidth]{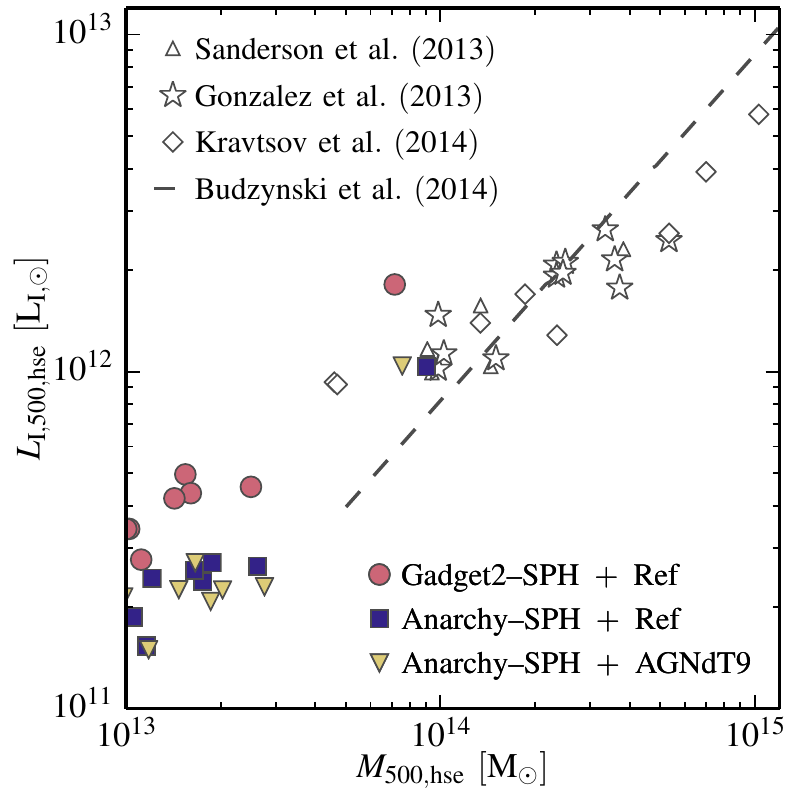}
\caption{I-Band luminosity within $R_{500,\rm{hse}}$ as a function of
  $M_{500,\rm{hse}}$ at $z = 0$ in the L050N0752 \anarchy SPH (blue squares) and
  \gadget SPH (red circles) simulations.  The yellow triangles show the haloes
  extracted from the \anarchy SPH run with an improved AGN model (AGNdT9).Data points
  correspond to the observations of \citet{Sanderson2013} (triangles),
  \citet{Gonzalez2013} (stars), \citet{Kravtsov2014} (diamonds) and the dashed line
  represents the SDSS image stacking results of \citet{Budzynski2014}. Where
  necessary, observations were converted to the I-band following \citet{LeBrun2014}.
  The \gadget SPH run overestimates the I-band luminosity in the group- and
  cluster-size objects as expected from the absence of passive galaxies in that
  simulation (see Fig.~\ref{fig:passive}).  }
\label{fig:Lopt_M500}
\end{figure}

The significant difference in star formation rates in massive haloes seen between the
two formulations of SPH can have consequences for quantities that are directly
observable. An example of such an observable is the $I$-band luminosity of groups and
clusters \citep[e.g.][]{Sanderson2013}. For galaxies with similar masses and
metallicities (as is the case when comparing matched pairs of galaxies extracted from
both our simulations), a higher $I$-band luminosity indicates a younger population of
stars and a higher star formation rate over the last billion years. In
Fig. \ref{fig:Lopt_M500}, we show the $I$-band luminosity as a function of halo mass
$M_{500,{\rm hse}}$. The values are computed by generating mock-observations of our
haloes as described by \cite{LeBrun2014}. Their procedure allows us to compute the
halo mass and radius assuming hydro-static equilibrium as is done in observations of
actual clusters. The (Cousin) $I$-band luminosity is computed within $R_{500,{\rm
    hse}}$, the overdensity radius inferred by assuming hydrostatic equilibrium in
the analysis of the mock observations. For comparison, we show observational data
taken from \cite{Sanderson2013}, \cite{Gonzalez2013} and \cite{Kravtsov2014} as well
as the SDSS image stacking result of \cite{Budzynski2014}. In all cases we selected
only clusters at $z<0.25$.

As expected from the previous analysis of the star formation rates, we find that the
$I$-band luminosity in the groups and clusters extracted from the simulation using
the \gadget flavour of SPH is higher than when using \anarchy.  It is also higher
than the trend extrapolated from observational data as expected from our analysis of
the specific star formation rates and the passive fractions for massive
($M_*>10^{11}\msun$) galaxies.  In the same figure, we also show the group and
cluster luminosities extracted from the simulation using the AGNdT9 model and the
\anarchy SPH scheme. The $I$-band luminosity as a function of mass for that model is
very similar to the one obtained using the Ref model. The differences between the
\gadget and \anarchy based simulations are much larger. However, as discussed
earlier, changing the model parameters for feedback from star formation will have an
even larger effect.

\subsection{ISM and CGM gas}

We now turn to the gas inside galaxies or in their direct vicinity. The state of this
gas will retain some of the properties of the IGrM but will also be directly affected
by the subgrid models.

\begin{figure}
\includegraphics[width=1.\columnwidth]{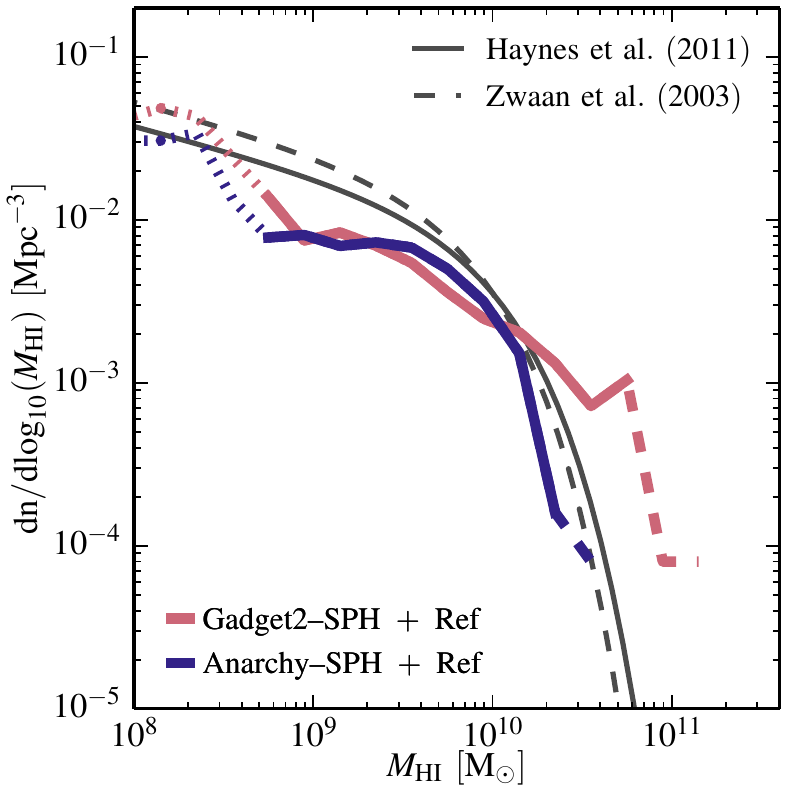}
\caption{The mass function of \HI-like gas (gas with $n_{\rm H} >
  0.01~\rm{cm}^{-3},~T<10^{4.5}~\rm{K}$) in the L050N0752 \anarchy SPH (blue line)
  and \gadget SPH (red line) simulations. Curves are drawn with dotted lines below a
  mass scale of 300 star particles, and with a dashed line style where sampled by fewer
  than 10 galaxies per $0.2~\rm{dex}$ mass bin. The solid and dashed grey lines show
  the best-fitting Schechter fits to the ALFALFA data by \citet{Haynes2011} and
  HIPASS data by \citet{Zwaan2003}, respectively. The simulation using the \gadget
  SPH formulation strongly overestimates the abundance of massive \HI clouds.}
\label{fig:HIMF}
\end{figure}

We first focus on the cold and dense phase of the gas. With the help of careful
simulations using radiative transfer, \cite{Rahmati2013} showed that cold
($T<10^{4.5}~\rm{K}$) and dense ($n_{\rm H} > 0.01~\rm{cm}^{-3}$) gas is a good
proxy for \HI gas. They provide a fitting function to compute \HI, but for the purpose
of this paper, setting the \HI fraction to $1$ for all this cold and dense gas is a
sufficiently good approximation. In Fig. \ref{fig:HIMF}, we show the mass function of the \HI
gas in the \anarchy (blue line) and \gadget (red line) simulations. We use dashed
lines when the mass bins contain fewer than $10$ objects and dotted lines when the
\HI mass corresponds to fewer than $300$ SPH particles.  We measured the \HI mass
using fixed spherical apertures placed at the centre of each subhalo in order to only
select the gas in the ISM and circum-galactic medium (CGM). As a point of reference,
we show the best-fitting Schechter functions to the data of \cite{Haynes2011}
(ALFALFA survey) and \cite{Zwaan2003} (HIPASS survey). 

As expected from the non-disruption of cold gas in the hot halo, there is an
over-abundance of massive \HI objects in the simulation using the \gadget variant of
SPH. Whilst the simulation using \anarchy is in reasonable agreement with the
observations, the same model using \gadget overshoots the break in the mass function
and vastly over-predicts the abundance of \HI clouds of mass $M_{\rm
  HI}>10^{10}~\msun$. Both simulations under-predict the abundance of low-mass
($M_{\rm HI}\lesssim 2\times10^9~\msun$) \HI clouds. As is shown by Crain et al. (in
prep) for \anarchy, this is a resolution effect. Simulations run with both flavours
of SPH exhibit the same behaviour in that regime and can then likely be rescued in a
similar way by increasing the resolution.

The discrepancy at the high-mass end is another sign that the densest gas clumps
found in the group- and cluster-like haloes are not disrupted by the hot halo. They
also seem to survive AGN activity and the effect of stellar feedback. These large
pools of cold gas in massive haloes are not observed and are likely to be responsible
for the spurious star formation seen in the largest galaxies (Figs. \ref{fig:SSFR}
and \ref{fig:passive}).  We note that it might be possible to modify the AGN subgrid
model so as to disrupt those clouds without breaking other constraints imposed on the
model. However, it seems unlikely that this purely numerical issue can be completely
alleviated. Furthermore, the abundance of spurious cold clumps will increase with the
resolution (as larger fluctuations in the density distribution can be sampled),
implying that the AGN activity needed to suppress them would also have to be modified.

\section{Summary \& Conclusion}
\label{sec:summary}

The aim of this study was to investigate the effects of the improved hydrodynamics
solver and time stepping used for the \eagle suite of cosmological simulations
\citep{Schaye2015,Crain2015}. By running the same simulation without re-calibrating
the subgrid model parameters with both \eagle's \anarchy and the standard \gadget
formulations of the SPH equations, we were able to isolate the effects of the
hydrodynamics solver. Thanks to the use of the pressure-entropy formulation of SPH
\citep{Hopkins2013}, a more stable kernel function \citep{Dehnen2012}, a small amount
of numerical diffusion \citep{Price2008}, an improved viscosity switch
\citep{Cullen2010} and the \cite{Durier2012} time-step limiter, the \anarchy flavour
of SPH is able to reproduce a large set of hydrodynamical tests more accurately than
the \gadget flavour \citep[Dalla Vecchia in prep.,][]{Sembolini2015}. Here we
investigated whether the better mixing of gas phases implied by these changes, as
well as the improved treatment of viscosity in shear flow, have consequences for the
simulation of haloes and galaxies. Our analysis of the differences can be summarized
as follows:

\begin{enumerate}

  \item Except for the most massive objects, the masses and sizes of the simulated
    galaxies are largely unaffected by the choice of SPH flavour. Uncertainties in
    the subgrid parameters lead to much larger differences (Figs. \ref{fig:GSMF} and
    \ref{fig:Sizes}).\\

  \item The absence of the \cite{Durier2012} time-step limiter leads to somewhat more
    efficient feedback, as expected from the non-conservation of energy occurring in
    feedback events when the limiter is neglected. For low-mass galaxies its effect
    is larger than that of the choice of hydro solver but small compared to the
    changes in the subgrid models for feedback (Figs. \ref{fig:no_limiter} and
    \ref{fig:Sizes}). For AGN feedback the time-step limiter might have a similar or
    stronger effect since the energy per feedback event is greater than for stellar
    feedback.\\

  \item The star formation rates of galaxies in small haloes, where the cooling time
    is smaller than the dynamical time, are unaffected by the change of hydrodynamics
    scheme. However, in massive haloes the star formation rates are much higher in
    the simulation using \gadget SPH (Figs. \ref{fig:SSFR}, \ref{fig:passive} and
    \ref{fig:Lopt_M500}). These differences in behaviour can be related to the lower
    quenching power of the AGN activity in that simulation. The lack of phase mixing,
    coming from the spurious artificial surface tension appearing at contact
    discontinuities, prevents cold dense gas from dissolving into the hot halo
    (Figs. \ref{fig:rhoT_LSS} and \ref{fig:maps}).\\

  \item This cold dense gas then reaches the central galaxies and leads to increased
    star formation (Figs. \ref{fig:SSFR} and \ref{fig:passive}) in both the central
    galaxies and intragroup medium (Fig. \ref{fig:M200_SFR}). This also leads to a
    lower hot gas fraction in the haloes (Fig. \ref{fig:gas_M500}) and an
    oversestimate of the \HI mass (Fig. \ref{fig:HIMF}).

\end{enumerate}

Our results indicate that the improved hydrodynamics scheme plays a significant role
in hot hydrostatic gas haloes, but not for lower-mass galaxies. Our results are
resolution dependent and it is possible that simulations performed at much higher
resolution will be more sensitive to the accuracy of the hydrodynamics
solver. Finally, we also stress that some of the differences between the simulations
could potentially be cancelled by changing the values of some of the subgrid
parameters.

\section*{Acknowledgements}
We thank the anonymous referee for their help improving this paper.  This work
would have not be possible without Lydia Heck and Peter Draper's technical
support and expertise.  This work was supported by the Science and Technology
Facilities Council (grant number ST/F001166/1); European Research Council (grant
numbers GA 267291 ``Cosmiway'' and GA 278594 ``GasAroundGalaxies'') and by the
Interuniversity Attraction Poles Programme initiated by the Belgian Science
Policy Office (AP P7/08 CHARM).  This work used the DiRAC Data Centric system at
Durham University, operated by the Institute for Computational Cosmology on
behalf of the STFC DiRAC HPC Facility (www.dirac.ac.uk). This equipment was
funded by BIS National E-infrastructure capital grant ST/K00042X/1, STFC capital
grant ST/H008519/1, and STFC DiRAC Operations grant ST/K003267/1 and Durham
University. DiRAC is part of the National E-Infrastructure.  We acknowledge
PRACE for awarding us access to the Curie machine based in France at TGCC, CEA,
Bruy\`eres-le-Ch\^atel. RAC is a Royal Society University Research Fellow.

\bibliographystyle{mnras} 
\bibliography{./bibliography.bib}

\label{lastpage}

\end{document}